\documentclass[11pt,a4paper]{article}
 \pdfoutput=1
\usepackage{jheppub}
\usepackage[american]{babel}
\usepackage{bbm}

\newcommand{\dvol}{d\mathrm{vol}}
\newcommand{\vol}{\mathrm{vol}}
\newcommand{\Vol}{\mathrm{Vol}}

\newcommand{\trans}{{\sf T}}

\newcommand{\parfrac}[2]{\frac{\partial #1}{\partial #2}}

\newcommand{\rep}[1]{\ensuremath{\mathbf{#1}}}

\newcommand{\wb}{\overline}
\newcommand{\wt}{\widetilde}

\newcommand{\pdag}{{\phantom{\dag}}}
\newcommand{\matht}[1]{$\boldsymbol{#1}$}

\newcommand{\ie}{\textit{i.e.}}

\numberwithin{equation}{section}

\newcommand{\Dslash}{D\!\!\!\!\slash\,}
\newcommand{\nn}{\nonumber}

\newcommand{\be}{\begin{equation}} \newcommand{\ee}{\end{equation}}
\newcommand{\bea}{\begin{equation} \begin{aligned}} \newcommand{\eea}{\end{aligned} \end{equation}}

\newcommand{\cA}{\mathcal{A}}
\newcommand{\cB}{\mathcal{B}}
\newcommand{\cC}{\mathcal{C}}
\newcommand{\cD}{\mathcal{D}}

\newcommand{\cG}{\mathcal{G}}
\newcommand{\cH}{\mathcal{H}}

\newcommand{\cL}{\mathcal{L}}
\newcommand{\cM}{\mathcal{M}}
\newcommand{\cN}{\mathcal{N}}
\newcommand{\cO}{\mathcal{O}}
\newcommand{\cP}{\mathcal{P}}
\newcommand{\cQ}{\mathcal{Q}}
\newcommand{\cR}{\mathcal{R}}

\newcommand{\cV}{\mathcal{V}}
\newcommand{\cW}{\mathcal{W}}

\newcommand{\cZ}{\mathcal{Z}}

\newcommand{\bC}{\mathbb{C}}

\newcommand{\bP}{\mathbb{P}}

\newcommand{\bR}{\mathbb{R}}
\newcommand{\bZ}{\mathbb{Z}}

\newcommand{\fh}{\mathfrak{h}}
\newcommand{\fm}{\mathfrak{m}}
\newcommand{\fM}{\mathfrak{M}}
\newcommand{\fn}{\mathfrak{n}}

\newcommand{\fR}{\mathfrak{R}}

\newcommand{\ft}{\mathfrak{t}}

\newcommand{\sQ}{\mathsf{Q}}

\def\su{\mathfrak{su}}

\DeclareMathOperator{\Tr}{Tr}

\DeclareMathOperator{\sign}{sign}
\DeclareMathOperator{\rank}{rank}
\DeclareMathOperator{\re}{\mathbb{R}e}
\DeclareMathOperator{\im}{\mathbb{I}m}
\DeclareMathOperator*{\Res}{Res}
\DeclareMathOperator*{\JKres}{JK-Res}

\title{Supersymmetric partition functions on Riemann surfaces}

\author[a,b]{Francesco Benini}
\author[c,d]{and Alberto Zaffaroni}

\affiliation[a]{International School for Advanced Studies (SISSA), \\
via Bonomea 265, 34136 Trieste, Italy --- INFN, Sezione di Trieste}
\affiliation[b]{Blackett Laboratory, Imperial College London, \\
South Kensington Campus, London SW7 2AZ, United Kingdom}
\affiliation[c]{Dipartimento di Fisica, Universit\`a di Milano-Bicocca, \\
I-20126 Milano, Italy}
\affiliation[d]{INFN, sezione di Milano-Bicocca, I-20126 Milano, Italy}

\emailAdd{fbenini@sissa.it}
\emailAdd{alberto.zaffaroni@mib.infn.it}

\preprint{SISSA 28/2016/FISI}

\abstract{We present a compact formula for the supersymmetric partition function of 2d $\cN=(2,2)$, 3d $\cN=2$ and 4d $\cN=1$ gauge theories on $\Sigma_g \times T^n$ with partial topological twist on $\Sigma_g$, where $\Sigma_g$ is a Riemann surface of arbitrary genus and $T^n$ is a torus with $n=0,1,2$, respectively. In 2d we also include certain local operator insertions, and in 3d we include Wilson line operator insertions along $S^1$. For genus $g=1$, the formula computes the Witten index. We present a few simple Abelian and non-Abelian examples, including new tests of non-perturbative dualities. We also show that the large $N$ partition function of ABJM theory on $\Sigma_g \times S^1$ reproduces the Bekenstein-Hawking entropy of BPS black holes in AdS$_4$ whose horizon has $\Sigma_g$ topology.}

\begin{document}

\setcounter{tocdepth}{2}
\maketitle

%
%

\section{Introduction}

Supersymmetric partition functions (\ie{} path-integrals) of quantum field theories on Euclidean compact manifolds are an extremely powerful tool to study non-perturbative properties of those theories. On the one hand, as one varies the compact manifold where the theories are placed and the supersymmetric background in which they are immersed (\ie{} the supersymmetric sources turned on), one gets access to a big deal of physical information such as correlation functions of operators and spectra of operators and states. On the other hand, keeping some supersymmetry in the process allows one to apply techniques, such as supersymmetric localization \cite{Witten:1988ze, Pestun:2007rz}, to explicitly and exactly compute those partition functions. This makes the program quantitative, not just qualitative.

In this note we study the partition function of two, three and four-dimensional supersymmetric gauge theories (with $\cN=(2,2)$, $\cN=2$ and $\cN=1$ supersymmetry, respectively) on a Riemann surface $\Sigma_g$ of genus $g$ times a torus $T^n$ (with $n=0,1,2$ respectively). In order to preserve (half of the) supersymmetry, we simply perform topological twist%
\footnote{In 2d, or upon reduction to 2d, we perform A-type topological twist on $\Sigma_g$.}
on $\Sigma_g$, \ie{} we turn on a background vector field coupled to the R-symmetry, equal and opposite to the spin connection. A technical assumption is then that the theories have a non-anomalous $U(1)_R$ R-symmetry. In fact this note is the natural generalization of \cite{Benini:2015noa}---where the case of the sphere, $g=0$, was studied---to higher genus. We will often refer to \cite{Benini:2015noa} for further details.

The partition functions become particularly interesting when the theories have some flavor symmetry. In that case we can turn on a background for the bosonic fields in external vector multiplets coupled to the flavor symmetry. It turns out that we can turn on a (quantized) magnetic flux on $\Sigma_g$ and a complex ``fugacity'' (whose definition depends on the dimension) along the Cartan subalgebra of the flavor group.%
\footnote{One could also turn on flat connections on $\Sigma_g$, but they do not affect the answer.}
We parametrize them by
\be
\fn = \frac1{2\pi} \int_{\Sigma_g} F^\text{flav} \qquad\text{ and }\qquad v \;.
\ee
In 2d $v = \sigma^\text{flav}$, the complex scalar in the vector multiplet that gives rise to twisted masses; in 3d $v \simeq A_t^\text{flav} + i \sigma^\text{flav}$ where $A_t^\text{flav}$ is a flat connection along $S^1$ and $\sigma^\text{flav}$ is a real scalar (that gives real masses); in 4d $v \sim A_{\bar z}^\text{flav}$ a flat connection on $T^2$. The partition function is a function of these parameters, meromorphic in $v$.

We can also decorate the partition functions by the inclusion of certain operators. In the 2d case we can make arbitrary insertions of local twisted chiral operators $\Sigma(x)$. In 3d we can make arbitrary insertions of Wilson line operators $\cW_\cR[\gamma]$ that wrap $S^1$ at a fixed position on $\Sigma_g$. Probably in 4d one could easily add surface operators wrapping $T^2$, but we do not consider them here.

Notice that in 2d we compute the standard A-twisted partition function \cite{Witten:1988xj} of gauge theories. In 3d we more precisely compute a Witten index
\be
Z_\text{3d} = \Tr_\cH\, (-1)^F \, e^{-\beta H} \, e^{i A^\text{flav} J^\text{flav}} \;.
\ee
Here $H$ is the Hamiltonian of the theory on $\Sigma_g$, with the prescribed R- and flavor symmetry fluxes, and the real masses generated by $\sigma^\text{flav}$. Then $\cH$ is the Hilbert space of states of $H$, and $J^\text{flav}$ are the flavor charge operators. From the supersymmetry algebra and the standard Witten index argument, one obtains that the states contributing to the index are those with
\be
0 = \cQ^2 = H - \sigma^\text{flav} J^\text{flav} \;.
\ee
We sometimes call $Z_\text{3d}$ as the ``higher-genus topologically twisted index''.
The 4d partition function could be put in the form of an index as well.

We compute the partition functions with localization techniques. The computation parallels the one in \cite{Benini:2015noa} (see also \cite{Closset:2015rna}) for the case $g=0$, which in turn adapts the computation of the elliptic genus in \cite{Benini:2013nda, Benini:2013xpa} to the present situation and reduces the path-integral to the evaluation of Jeffrey-Kirwan residues \cite{JeffreyKirwan}. The novelty for $g>0$ is that there are more bosonic and fermionic zero-modes. After taking them into account one can put the result in the following schematic form:
\be
Z_{\Sigma_g \times T^n} = \sum_{u = u_{(\alpha)}} Z_\text{cl,1l} \big|_{\fm=0} \; \bigg( \det_{ab} \frac{\partial B_a}{\partial u_b} \bigg)^{g-1} \;,
\ee
where $Z_\text{cl,1l}$ (classical and one-loop contribution) is a function of the flavor parameters $(v,\fn)$ and similar gauge parameters $(u,\fm)$, obtained from the quadratic expansion of the action around BPS configurations, then
\be
iB_a = \frac{\partial \log Z_\text{cl,1l}}{\partial \fm_a} \;,
\ee
and $u_{(\alpha)}$ are a set of solutions to the equations, named Bethe Ansatz Equations (BAEs),
\be
e^{iB_a} = 1
\ee
such that a Vandermonde determinant is not zero. This expression has appeared before, for instance in \cite{Okuda:2012nx, Okuda:2013fea, Nekrasov:2014xaa, Gukov:2015sna, Okuda:2015yea} (where the name BAEs was given). We give a derivation of the formula within the Jeffrey-Kirwan residue framework, from which one can extract a precise prescription for what solutions to the BAEs should be kept, even in theories with a complicated matter content.%
\footnote{The formulation as Jeffrey-Kirwan residues is more general, while the expression given above is only valid if the roots of the BAEs are simple.}
To the best of our knowledge, the 4d case has not appeared before.

We present a few simple examples and applications of the formula in various dimensions, comparing with known results when they are available. In three dimensions we perform new tests of non-perturbative dualities, in particular of Aharony duality \cite{Aharony:1997gp} and the so-called ``duality appetizer'' \cite{Jafferis:2011ns}. The higher-genus index proves to be useful to identify topological sectors involved in the dualities. In four dimensions we perform a simple check of Seiberg duality \cite{Seiberg:1994pq}. We leave a more detailed analysis for the future.

Specialized to $g=1$ and turning off the flavor fluxes, $\fn=0$, the partition function becomes independent of the flavor fugacities $v$ and it computes the Witten index \cite{Witten:1982df}. This result should be taken with care, though. First, our computation is only valid when there is a continuous non-anomalous R-symmetry, even at $g=1$. Thus, in particular, we cannot use our formula to reproduce the Witten index of 4d super-Yang-Mills (SYM). Second, our formula computes the Witten index of the theory with generic fugacities $v$, which might be different from the index of the theory with $v=0$.%
\footnote{This is because we treat the superpotential as a $\cQ$-exact deformation, but one should be careful in doing that.}
With this proviso, we reproduce the Witten indices computed in \cite{Intriligator:2013lca} in the cases we consider.

One of the other purposes of this work is to generalize the counting of micro-states of a class of black holes in AdS$_4$ performed in \cite{Benini:2015eyy}.   
We then compute the higher-genus index of the three-dimensional ABJM theory \cite{Aharony:2008ug} in the large $N$ limit at leading order. We will use the large $N$ solution
of the BAEs found in \cite{Benini:2015eyy} to evaluate the index and its dependence on $g$. It was shown in \cite{Benini:2015eyy} that the $g=0$ twisted index reproduces the Bekenstein-Hawking entropy of certain BPS black holes in AdS$_4$; we show here that the $g>0$ index correctly reproduces the entropy of BPS black holes whose horizon has the topology of $\Sigma_g$.

The note is organized as follows. In section \ref{sec: 3d} we derive the formula of the higher-genus topologically twisted index of 3d theories. In sections \ref{sec: 2d} and \ref{sec: 4d} we give the generalization to 2d and 4d. In section \ref{sec: examples} we present various examples. In section \ref{sec: entropy} we compare the large $N$ index with the black hole entropy.

\

We have coordinated the release of this note with \cite{Closset:2016arn}, which has overlap with the material presented here.


\section{3d theories on \matht{\Sigma_g \times S^1}}
\label{sec: 3d}

We consider three-dimensional $\cN=2$ gauge theories with a $U(1)_R$ R-symmetry on $\Sigma_g \times S^1$, where $\Sigma_g$ is an arbitrary Riemann surface of genus $g$. This setup is the higher-genus generalization of the ``topologically twisted index'' studied in \cite{Benini:2015noa}. On $\Sigma_g$ we take vielbein $e^{1,2}$ and on $S^1$ we take $e^3 = \beta\, dt$ with $t \cong t+1$. In order to preserve some supersymmetry, we perform a partial topological twist on $\Sigma_g$, \ie{} we turn on a background connection $V_\mu$ on $\Sigma_g$ coupled to the R-symmetry current, such that it cancels the spin connection for half of the supercharges:
\be
V = - \frac12 \omega^{12} \;,\qquad\qquad W = dV = - \frac{R_s}4 \, e^1 \wedge e^2 \;,\qquad\qquad \frac1{2\pi} \int_{\Sigma_g} W = g-1 \;.
\ee
Here $\omega_\mu^{ab}$ is the spin connection, while $R_s$ is the scalar curvature on $\Sigma_g$. Since the BRST-invariant sector of the twisted theory becomes independent of the metric on $\Sigma_g$ \cite{Witten:1988xj}, we will not need to specify it.

The supersymmetry parameters $\epsilon$, $\tilde\epsilon$ are commuting spinors, have R-charge $-1$, and solve the twisted Killing spinor equation
\be
D_\mu \epsilon = D_\mu \tilde\epsilon = 0 \;.
\ee
Since $D_\mu \epsilon = (\partial_\mu - i \omega_\mu^{12} P_- )\epsilon$ (and similarly for $\tilde\epsilon$) where $P_\pm = (1\pm \gamma_3)/2$ are the chiral projectors on $\Sigma_g$, it follows that $\epsilon_+$, $\tilde\epsilon_+$ are constant while $\epsilon_- = \tilde\epsilon_- = 0$. The only exception is the flat torus, namely $g=1$, in which case also constant $\epsilon_-$, $\tilde\epsilon_-$ are solutions: obviously $T^2$ does not break any supersymmetry. However as soon as we turn on flavor fugacities or fluxes, or we take a curved metric on $T^2$, those supercharges get broken.

We are interested in gauge theories of vector and chiral multiplets. They are specified by a gauge group $G$, a matter representation $\fR$ of $G$ for the chiral multiplets $\Phi$, and a superpotential $W(\Phi)$ which is a holomorphic function, homogeneous of degree $2$ with respect to a choice of R-charges $q_\phi$. The choice of R-charges controls the coupling to the R-symmetry background, moreover the presence of a net R-symmetry flux requires the choice to satisfy the quantization condition
\be
\label{quantization R-charges}
q_\cO \, (g-1) \in \bZ
\ee
for all gauge-invariant operators $\cO$. In three dimensions, besides the super-Yang-Mills Lagrangian we can also add a supersymmetric Chern-Simons (CS) term, which is a quantized Killing form $k$ for $G$: the parameters are one integer for each simple factor, and an integer matrix for the Abelian factors. Thus, the Lagrangian of the theory is
\be
\cL = \cL_\text{YM} + \cL_\text{CS} + \cL_\text{mat} + \cL_W \;.
\ee
The various terms (spelled out in appendix \ref{app: notation}) are essentially equal to those on flat space, the only difference being that they are coupled to the background.

The supersymmetry variations in our notation are given in appendix \ref{app: notation}. They realize the superalgebra $\su(1|1)$, whose bosonic subalgebra $\mathfrak{u}(1)$ generates rotations of $S^1$ mixed with gauge and flavor rotations:
\be
\label{susy algebra}
\{Q, \wt Q\} = -i \cL^A_v - \delta_\text{gauge}(\tilde \epsilon^\dag\epsilon\, \sigma) \;,\qquad\qquad Q^2 = \wt Q^2 = 0 \;,\qquad\qquad v^\mu = \tilde\epsilon^\dag \gamma^\mu \epsilon \;.
\ee
Here $\cL^A_v$ is the gauge-covariant Lie derivative (including the R-symmetry connection) along the covariantly constant (and Killing) vector field $v^\mu$.%
\footnote{The explicit expression of the Lie derivative of fields of various spins can be found in appendix B.1 of \cite{Benini:2013yva}.}
In fact $v = \beta^{-1} \tilde\epsilon^\dag\epsilon \, \partial_t$.

The supercharges are compatible with a non-trivial background for the flavor symmetries, \ie{} background values for external vector multiplets coupled to the flavor currents (as in \cite{Benini:2015noa}). In three dimensions, flavor symmetries include the topological symmetries generated by $J_\mu^\text{T} = (* F)_\mu$ for each Abelian gauge field. It turns out that, up to flavor rotations, one can turn on a constant flat connection $A_3^\text{flav}$ along $S^1$, a constant $\sigma^\text{flav}$, and a flux $F_{12}^\text{flav}$ on $\Sigma_g$ accompanied by
\be
D^\text{flav} = i F_{12}^\text{flav} \;,
\ee
all taking values in the Cartan subalgebra of the flavor symmetry group $G_F$. It is easy to check from \eqref{vector variations} that the supersymmetry conditions $Q\lambda^\text{flav} = \wt Q \lambda^\text{flav} = 0$, and similarly for $\lambda^{\dag\,\text{flav}}$, are met. We parametrize the background by
\be
v = \beta \bigl( A_3^\text{flav} + i \sigma^\text{flav} \bigr) \;,\qquad\qquad \fn = \frac1{2\pi} \int_{\Sigma_g} F^\text{flav} \;.
\ee
The ``complexified flat connection'' $v$ actually takes values on the complexified maximal torus of $G_F$, and the flux $\fn$ is GNO quantized (meaning that $\gamma(\fn) \in \bZ$ for the weights $\gamma$ of all representations of $G_F$, \ie{} $\fn$ is in the coroot lattice). These parameters make the partition function on $\Sigma_g \times S^1$ an interesting function.

We can also include Wilson line operators $\cW_\cR$, where $\cR$ are representations of $G$, along $S^1$ and sitting at arbitrary points on $\Sigma_g$. They are constructed in the standard way as
\be
\cW_\cR = \Tr_\cR \text{Pexp} \oint_{S^1} \big( i A - \sigma \, e^3 \big) \;,
\ee
where Pexp is the path-ordered exponential.

For genus $g>0$, one could turn on flavor flat connections on $\Sigma_g$ as well. Since such connections do not show up in the supersymmetry algebra (\ref{susy algebra})---\ie{} they are not central charges---they do not affect the partition function and we will not introduce them.

The quantity of interest, that we can call a ``higher-genus topologically twisted index'', is the partition function of the theory, possibly with Wilson line insertions, on $\Sigma_g \times S^1$:
\be
\label{def of partition function}
Z_{\Sigma_g \times S^1}(t_j, \cR_\alpha) = \int \cD\varphi\, \bigg( \prod\nolimits_\alpha \cW_{\cR_\alpha} \bigg) \, e^{-S[\varphi, t_j]} \;,
\ee
where $t_j$ are collectively the parameters of the background.

\subsection{The derivation of the formula}
\label{sec: derivation}

We wish to compute the partition function (\ref{def of partition function}) on $\Sigma_g \times S^1$ with localization techniques \cite{Witten:1988xj, Witten:1991zz}. In doing the computation, one encounters a system of bosonic and fermionic zero-modes that can be dealt with using the method in \cite{Benini:2013nda, Benini:2013xpa} (see also \cite{Hori:2014tda, Benini:2015noa, Closset:2015rna, Closset:2015ohf}), originally envisaged to compute the elliptic genus of two-dimensional gauge theories. In fact, one can follow almost verbatim the computation in \cite{Benini:2015noa}, with a few modification that we will discuss here.

We perform localization with respect to the supercharge $\cQ = Q + \wt Q$ where one takes $\epsilon = \tilde\epsilon$. The BPS equations give a moduli space $\wt \cM_\text{BPS}$ of complexified BPS configurations modulo gauge transformations
\be
\wt \cM_\text{BPS} = \Big\{ D = i F_{12} \,,\, F_{13} = F_{23} = 0 \,,\, D_\mu \sigma = 0 \Big\} / \cG \;.
\ee
As argued in \cite{Benini:2015noa}, only configurations with constant $D$ contribute to the real path-integral, and we call $\cM_\text{BPS}$ such a moduli space. For $g=0$, $\cM_\text{BPS}$ is parametrized by ``complexified flat gauge connections'' $u$ and gauge fluxes $\fm$,
\be
u = \beta (A_3 + i \sigma) \;,\qquad\qquad \fm = \frac1{2\pi} \int_{\Sigma_g} F \;,
\ee
where $\re u$ is along the maximal torus $H$ of $G$, $\im u$ and $\fm$ are along the Cartan subalgebra $\fh$, $\fm$ is GNO quantized to the coroot lattice $\Gamma_\fh$, and we mod by Weyl transformations. In other words, $\cM_\text{BPS}^{g=0} = (H \times \fh \times \Gamma_\fh)/W$. The novelty for $g>0$ is that we also have flat connections on $\Sigma_g$. For generic $u$, the flat connections on $\Sigma_g$ should be in the maximal torus $H$, giving an extra factor $H^{2g}$. Along special complex-codimension-1 hyperplanes in the $u$ plane where the commutant of $u$ in $G$ is non-Abelian, one can have non-Abelian flat connections on $\Sigma_g$ and their space is larger. As we will see, those hyperplanes do not contribute to the path-integral and can be removed. This is similar to the behavior of chiral multiplets already observed for $g=0$. For generic values of $u$, the BPS equations set the chiral multiplets to zero; along special hyperplanes there can be non-trivial solutions, however those hyperplanes do not contribute.

Thus, for generic values of $u$, the bosonic zero-modes take values on
\be
\fM \times H^{2g} \qquad\text{ where }\qquad \fM = H \times \fh \;.
\ee
There are also fermionic zero-modes, from the Cartan gaugini. It turns out that they form 0d off-shell supermultiplets, which can be thought of as the dimensional reduction of 2d $\cN{=}(0,2)$ supermultiplets.

The fermionic zero-modes are constant on $S^1$ and satisfy $\Dslash_{(\Sigma)} \lambda = 0$ on the Riemann surface. In components this is $\partial_{\bar z} \lambda_+ = 0$ and $(\partial_z - i \omega_z^{12}) \lambda_- = 0$, where $e^z = e^1 + i e^2$ is the holomorphic vielbein, in particular $\lambda_+$ is a constant (the unique holomorphic function) and $\lambda_-^\dag$ is a holomorphic differential. Therefore for any value of $g$ there is a 0d vector multiplet---comprised of $(u, \bar u, \lambda_0^\pdag, \lambda_0^\dag, D_0)$ where $u$, $\bar u$, $D_0$ are bosonic and $\lambda_0^\pdag$, $\lambda_0^\dag$ are fermionic---that parametrizes the constant zero-modes. Besides $u$, $\bar u$, the other ones are
\be
\lambda_0^\pdag = \beta\, \tilde\epsilon^\dag \lambda \;,\qquad\qquad \lambda_0^\dag = \beta\, \lambda^\dag \epsilon \;,\qquad\qquad D_0 = \beta\, \tilde\epsilon^\dag\epsilon \, (D - i F_{12})
\ee
where we implicitly took the constant mode in each case. We get the supersymmetry algebra
\bea
\label{variations zero-modes vector}
Qu&=0 \qquad& Q \bar u &= i\lambda_0^\dag \qquad& Q \lambda_0^\pdag &= - D_0 \qquad& Q\lambda_0^\dag &= 0 \qquad& QD_0 &= 0 \\
\wt Qu &= 0 \qquad& \wt Q\bar u &= i\lambda_0^\pdag \qquad& \wt Q \lambda_0^\pdag &= 0 \qquad& \wt Q \lambda_0^\dag &= D_0 \qquad& \wt Q D_0 &= 0 \;.
\eea
The mode $D_0$ is an auxiliary zero-mode.
Notice that $Q^2 = \wt Q^2 = \{Q,\wt Q\} = 0$ on the zero-mode subspace, since the zero-modes are translationally invariant and commute with $\sigma$.

For $g>0$ there are $g$ 0d chiral multiplets---comprised of $\big( a^{(\alpha)}, \bar a^{(\alpha)}, \eta_0^{(\alpha)}, \eta_0^{(\alpha)\dag} \big)$ where the first two are bosonic and the last two fermionic---that parametrize the modes proportional to the $g$ holomorphic differentials $\omega_\alpha$ on $\Sigma_g$. It is convenient to define the following 1-forms on $\Sigma_g$:
\be
\eta_j = \frac i2 \tilde \epsilon^\dag \gamma_j \lambda \qquad\text{with}\qquad \eta_z = 0 \;,\qquad\qquad
\eta_j^\dag = \frac i2 \lambda^\dag \gamma_j \epsilon \qquad\text{with}\qquad \eta^\dag_{\bar z} = 0
\ee
where $j=1,2$ only. Since $\tilde\epsilon^\dag \gamma_{\bar z}\lambda = \tilde\epsilon_+^\dag \lambda_-$ and $\lambda^\dag \gamma_z \epsilon = \lambda_-^\dag \epsilon_+$, those forms contain the negative-chirality zero-modes in $\lambda$. Using that all modes are independent from $t$ and along the Cartan subalgebra, from (\ref{vector variations}) we obtain the transformations
\bea
\label{variations zero-modes on Sigma}
Q A_j &= \eta^\dag_j \;,\qquad\qquad& Q \eta^\dag_j &= 0 \;,\qquad\qquad& Q \eta_{\bar z} &= i \tfrac{\tilde\epsilon^\dag \epsilon}\beta D_{\bar z} u \\
\wt Q A_j &= \eta_j \;, & \wt Q \eta_j &= 0 \;, & \wt Q \eta^\dag_z &= i \tfrac{\tilde\epsilon^\dag \epsilon}\beta D_z u \;.
\eea
We can parametrize the flat connections on $\Sigma$ as
\be
A_{(\Sigma)} = \sum_{\alpha=1}^g a^{(\alpha)} \, \omega_\alpha + A_z^\text{ref} dz + \sum_{\alpha=1}^g \bar a^{(\alpha)} \, \bar\omega_\alpha + A_{\bar z}^\text{ref} d\bar z \;,
\ee
where $A^\text{ref}$ is a reference background connection such that $dA^\text{ref} = 2\pi \fm \, \dvol_{\Sigma_g} / \Vol(\Sigma_g)$. Similarly we parametrize
\be
\eta^\dag = \sum\nolimits_\alpha \eta_0^{(\alpha)\dag} \omega_\alpha \;,\qquad\qquad \eta = \sum\nolimits_\alpha \eta_0^{(\alpha)} \bar\omega_\alpha \;.
\ee
We thus obtain the superalgebra
\bea
Q a^{(\alpha)} &= \eta_0^{(\alpha)\dag} \;,\qquad& Q \bar a^{(\alpha)} &= 0 \;,\qquad& Q \eta_0^{(\alpha)} &= Q \eta_0^{(\alpha)^\dag} = 0 \\
\wt Q a^{(\alpha)} &= 0 \;,\qquad& \wt Q \bar a^{(\alpha)} &= \eta_0^{(\alpha)} \;,\qquad& \wt Q \eta_0^{(\alpha)} &= \wt Q \eta_0^{(\alpha)^\dag} = 0 \;.
\eea

The on-shell classical action has been computed in \cite{Benini:2015noa}. Let us distinguish the background fields $(v,\fn)$ for ``standard'' flavor symmetries from those $(w,\ft)$ for the topological symmetries, as they appear differently in the Lagrangian. We use the notation $y = e^{iv}$ to mean that $y^\gamma = e^{i\gamma(v)}$ for the flavor weights $\gamma$, and similarly $\xi = e^{iw}$. The CS action contributes $Z_\text{CS} = e^{i k(u,\fm)}$, where $k$ is the CS Killing form. In particular each simple factor $G_I$ in $G$ gives
\be
\label{Z CS n-Ab}
Z_\text{CS}^\text{n-Ab} = x^{k\fm} = \prod_{i=1}^{\rank G_I} x_i^{k_I \fm_i} \;,
\ee
while the Abelian factors in $G$ give
\be
Z_\text{CS}^\text{Ab} = \prod\nolimits_{i,j} x_i^{k_{ij} \fm_j} \;.
\ee
These expressions are valid, with the substitution $(u,\fm) \to (v,\fn)$ or $(u,\fm) \to (w,\ft)$, for gauge-flavor and flavor-flavor CS terms as well. Gauge-R or flavor-R CS terms give
\be
Z_\text{CS}^\text{R} = \prod\nolimits_i x_i^{(g-1) k_{iR}} \;,
\ee
since the R-symmetry flux is fixed by supersymmetry. The coupling of topological currents $J_\mu^\text{T}$ to background vector multiplets leads to a factor
\be
Z_\text{CS}^\text{T} = x^\ft \xi^\fm
\ee
for each topological symmetry. Finally, each Wilson line operator insertion gives a factor
\be
\cW_\cR = \Tr_\cR x = \sum\nolimits_{\rho\in\cR} x^\rho
\ee
which is the character of $\cR$.

The one-loop determinants of small quadratic fluctuations around the BPS configurations have been computed in \cite{Ohta:2012ev, Okuda:2013fea, Gukov:2015sna} (see also \cite{Benini:2015noa}). From chiral multiplets we get
\be
\label{Z 1-loop chiral}
Z_\text{1-loop}^\text{chiral} = \prod_{\rho\in\fR} \Big( \frac{x^{\rho/2} y^{\gamma/2} }{ 1-x^\rho y^\gamma } \Big)^{\rho(\fm) + \gamma(\fn) + (g-1)(q_\rho -1)} \;,
\ee
where the product is over the weights of the gauge representation $\fR$. From vector multiplets we get
\be
\label{Z 1-loop gauge}
Z_\text{1-loop}^\text{gauge} = \prod_{\alpha\in G} (1-x^\alpha)^{1-g} (i\, du)^{\rank G} \;,
\ee
where the product is over the roots of $G$. $Z^\text{gauge}_\text{1-loop}$ is a holomorphic top-form on $\fM$. We will compactly call
$$
Z_\text{cl,1l}(x,\fm,y,\fn,\xi,\ft)
$$
the product of all classical (including the Wilson line operators) and one-loop contributions: this is a holomorphic top-form on $\fM$. Note that there is no dependence on the flat connections on $\Sigma_g$.

At this point we should integrate over all zero-modes. The integral over the modes $a^{(\alpha)}$, $\bar a^{(\alpha)}$ is trivial because $Z_\text{cl,1l}$ does not depend on them. They parametrize the space $H^{2g}$ of Abelian flat connections on $\Sigma_g$, therefore the integral gives a volume factor.%
\footnote{Since the final result is proportional to a normalization constant that we need to fix in one known example, say $U(1)$ Chern-Simons theory, we need not compute the volume factor here.}

Let us denote by $\cZ(u, \bar u, \lambda_0^\pdag, \lambda_0^\dag, D_0, a, \bar a, \eta_0^\pdag, \eta_0^\dag; \fm)$ the effective partition function of the zero-modes, obtained by integrating out all non-zero-modes in a given flux sector $\fm$. Setting to zero all fermionic and the auxiliary mode $D_0$, it coincides with the classical and one-loop contribution we have computed above:
\be
\cZ \Big|_{\lambda_0^\pdag = \lambda_0^\dag = \eta_0^\pdag = \eta_0^\dag = D_0 = 0} = Z_\text{cl,1l}(u, \bar u;\fm) \;.
\ee
The integral over fermionic zero-modes is a derivative, that we need to compute:
\be
\int d\lambda_0^\pdag d\lambda_0^\dag \bigg( \prod_{\alpha=1}^g d\eta_0^{(\alpha)} d\eta_0^{(\alpha)\dag} \bigg) \, \cZ = \bigg( \prod_{\alpha=1}^g \frac{\partial^2}{\partial \eta_0^{(\alpha)} \partial \eta_0^{(\alpha)\dag}} \bigg) \frac{\partial^2}{\partial\lambda_0^\pdag \partial\lambda_0^\dag} \, \cZ \;.
\ee
Such a derivative is fixed by supersymmetry, using some tricks in \cite{Gerasimov:2006zt, Okuda:2013fea, Benini:2015noa, Closset:2015rna}. For the sake of clarity, let us consider the case that $\rank G=1$. Since the effective action is topological on $\Sigma_g$, it takes the generic form
\be
\cZ = \cA \, \exp \int_{\Sigma_g} \Big[ \cB\, F + c\,\cG \, \eta^\dag \wedge \eta \Big]
\ee
where $\cA$, $\cB$ and $\cG$ are functions of $u, \bar u, \lambda_0^\pdag, \lambda_0^\dag, D_0$ and the constant $c=\big( \frac{\tilde\epsilon^\dag\epsilon}{\beta} \big)^{-1}$ has be inserted for later convenience. In particular the logarithm of $\cZ$ is linear in $\fm$ (which is indeed the case for $Z_\text{cl,1l}$) and there is no dependence on the flat connections on $\Sigma_g$. From the fact that $\cZ$ is supersymmetric and using (\ref{variations zero-modes vector}) and (\ref{variations zero-modes on Sigma}):
\be
\label{variation effective action}
0 = Q \cZ = i \lambda_0^\dag \parfrac{\cZ}{\bar u} - D_0 \parfrac{\cZ}{\lambda_0} + \cZ \cdot \int_{\Sigma_g} \Big[ \cB \, d\eta^\dag - i \, \cG\, \eta^\dag \wedge du \Big] \;.
\ee
The only terms containing $\eta^\dag$ but not $\eta$ are in the integral. Integrating by parts we conclude
\be
\cG = \parfrac{\cB}{iu} \;.
\ee
We also have $\cB = \partial \log \cZ / \partial \fm$, therefore
\be
\cG = \frac{\partial^2 \log \cZ}{\partial iu \, \partial \fm} \;.
\ee
Expanding $\eta$, $\eta^\dag$ into the zero-modes and assuming a normalized basis of holomorphic differentials on $\Sigma_g$, we find
\be
\prod_{\alpha=1}^g \frac{\partial^2 \cZ}{\partial \eta_0^{(\alpha)} \partial \eta_0^{(\alpha)\dag}} = c^g\, \cG^g \, \cZ\big|_{\eta = \eta^\dag = 0} = c^g\, \Big( \frac{\partial^2 \log \cZ}{\partial iu \, \partial \fm} \Big)^g \cZ \Big|_{\eta = \eta^\dag = 0} \;.
\ee
The only terms containing $\lambda_0^\dag$ but not $\lambda_0$ are the first two in (\ref{variation effective action}), which should cancel each other. Taking a derivative with respect to $\lambda_0^\dag$ we get
\be
D_0 \parfrac{^2 \cZ}{\lambda_0^\pdag \partial \lambda_0^\dag} = - i \parfrac{\cZ}{\bar u} \Big|_{\lambda_0^\pdag = \lambda_0^\dag = 0} \;.
\ee
We thus find, up to an unimportant multiplicative constant:
\be
\label{absorption of zero-modes}
\bigg( \prod_{\alpha=1}^g \frac{\partial^2}{\partial \eta_0^{(\alpha)} \partial \eta_0^{(\alpha)\dag}} \bigg) \frac{\partial^2}{\partial\lambda_0^\pdag \partial\lambda_0^\dag} \, \cZ = - \frac i{D_0} \, \parfrac{}{\bar u} \, \Big( \frac{\partial^2 \log \cZ}{\partial iu \, \partial \fm} \Big)^g \cZ \Big|_{\lambda_0^\pdag = \lambda_0^\dag = \eta_0^\pdag = \eta_0^\dag = 0} \;.
\ee

Then we should integrate over the bosonic zero-modes $u$, $\bar u$, $D_0$. As in \cite{Benini:2015noa}, on the domain $\fM$ of $u, \bar u$ there are singular hyperplanes (in the $\rank G=1$ case, those are just points) $H_i$ where some chiral multiplet becomes massless:
\be
H_i = \big\{ u \in \fM \,\big|\, e^{i \rho_i(u) + i \gamma_i(v)} = 1 \big\} \;.
\ee
There are also hyperplanes where some W-boson becomes massless (whose wavefunction on $\Sigma_g$ is a holomorphic differential):
\be
H_\alpha = \big\{ u \in \fM \,\big|\, e^{i \alpha(u)} = 1 \big\} \;.
\ee
These are precisely the zeros of the Vandermonde determinant $\prod_{\alpha\in G} (1-x^\alpha)$. As explained in \cite{Benini:2015noa} (following \cite{Benini:2013nda}), using (\ref{absorption of zero-modes}) and Stoke's theorem, the integral over $\fM$ reduces to a Cauchy contour integral around the hyperplanes $H_i$, $H_\alpha$ and around the regions at infinity. Then, as a result of the integration over $D_0$, the contour integral is taken at $D_0 = 0$, and only certain residues are picked up. Which ones depends on the choice of an auxiliary parameter $\eta \in \fh^*$, and it is controlled by the Jeffrey-Kirwan residue \cite{JeffreyKirwan, Brion1999, Szenes:2004}. The computation that leads to (\ref{absorption of zero-modes}) is easily generalized to the higher-rank case, as well the reduction to a contour integral (this is more intricate, and we refer to \cite{Benini:2013xpa} and \cite{Benini:2015noa}). The result is
\be
\label{index formula higher rank}
Z_{\Sigma_g \times S^1} = \frac1{|W|} \sum_{\fm \in \Gamma_\fh} \sum_{u_* \in \fM_\text{sing}^*} \JKres_{u = u_*} \big( \sQ_{u_*}, \eta \big) \, \bigg( \det_{ab} \frac{\partial^2 \log Z_\text{cl,1l} }{ \partial iu_a\, \partial \fm_b } \bigg)^g Z_\text{cl,1l} + \text{bound. contrib.}
\ee
Here $|W|$ is the order of the Weyl group; $\fM_\text{sing}^*$ is the collection of singular points in $\fM$ where at least $\rank G$ linearly independent singular hyperplanes meet; $\sQ_{u_*}$ is the set of charge covectors in $\fh^*$ of the singular hyperplanes passing through $u_*$. The Jeffrey-Kirwan residue operation performs integration along selected middle-dimensional contours around each of the points $u_*$, based on the choice of $\eta$, that satisfy the following conditions:
\begin{equation}
\label{BLSVcondition}
\JKres_{u=0}(\sQ_*,\eta) \, \frac{d Q_{j_1}(u)}{Q_{j_1}(u)} \wedge \cdots \wedge \frac{d Q_{j_r}(u)}{Q_{j_r}(u)} =
\begin{cases}
\sign \det (Q_{j_1} \dots Q_{j_r}) & \text{if } \eta\in \text{Cone}(Q_{j_1} \ldots Q_{j_r}) \\
0 & \text{otherwise}
\end{cases}
\end{equation}
where $\text{Cone}$ denotes the cone spanned by the vectors in the argument. The boundary contributions are integrals of the same form along middle-dimensional contours around infinity, constructed in \cite{Benini:2015noa}. Fortunately, one can usually avoid to evaluate the boundary contours by commuting sum and integration in (\ref{index formula higher rank}): this brings all singular points in the interior of $\fM$. More details can be found in \cite{Benini:2015noa, Benini:2013xpa}.

In the $\rank G = 1$ case, the formula simplifies to
\be
\label{index formula rank 1}
Z_{\Sigma_g \times S^1} = \frac1{|W|} \sum_{\fm \in \Gamma_\fh} \; \sum_{x_* \in \fM_\text{sing}, 0, \infty} \; \JKres_{x=x_*} \big( \sQ(x_*), \eta\big) \, \bigg( \frac{\partial^2 \log Z_\text{cl,1l} }{ \partial \log x\, \partial \fm} \bigg)^g Z_\text{cl,1l} \;,
\ee
where we mapped the cylinder to the punctured Riemann sphere by $x=e^{iu}$. The boundary contributions are captured by the very same JK residue, assigning to the points $x=0,\infty$ charges proportional to the effective CS coupling at infinity on the Coulomb branch:
\be
k_\pm \equiv k_\text{eff}(u=\pm\infty) = k + \frac12 \sum_i Q_i^2 \sign(\pm Q_i) \;,\qquad Q_0 = - k_+ \;,\qquad Q_\infty = k_- \;.
\ee
Here $Q_i$ are the charges of chiral multiplets if $G=U(1)$, and the weights if $G=SU(2)$. In particular, choosing $\eta>0$ one picks the residues from poles with associated positive charge; choosing $\eta<0$ one picks minus the residues from poles with associated negative charge. The two choices give the same answer.

For Abelian gauge groups these formul\ae{} give the final answer. In the non-Abelian case, instead, the formul\ae{} are not yet complete. One of the assumptions in the derivation of the JK residue in \cite{Benini:2013nda, Benini:2013xpa, Benini:2015noa} is that at each point $u_* \in \fM_\text{sing}^*$, the singular hyperplane arrangement is ``projective''. In the rank-one case this means that all fields that become massless at $u_*$ have charges with the same sign; in the higher-rank case this means that the weights $\rho$ of the massless fields at $u_*$ lie in a half-space. This condition is not met by the hyperplanes $H_\alpha$ from W-bosons, since for each $H_\alpha$ there is a coincident $H_{-\alpha}$. In order to correctly evaluate the contribution from the W-bosons, we need to resolve their singularities.

Among the various possibilities, we choose the following regularization of the vector multiplet one-loop determinant. Such determinant is equal to the inverse of the determinant for chiral multiplets of R-charge $0$ transforming as the roots $\alpha$ of $G$.%
\footnote{In \cite{Benini:2015noa} we have argued that the one-loop determinants suffer from a sign ambiguity, which shows up when there are Abelian factors in $G$ and it can be reabsorbed by a sign redefinition of $\xi$. Hence, to make our formul\ae{} lighter, we have included a factor $(-1)^{2\delta(\fm)}$ into the vector multiplet determinant, where $\delta = \frac12 \sum_{\alpha>0} \alpha$ is the Weyl vector.}
This follows from the Higgs mechanism. Suppose we add to the theory an adjoint chiral multiplet of R-charge $0$. Its diagonal components can get a VEV without breaking the R-symmetry. A W-boson along a root $\alpha$ combines with the component $\alpha$ of the chiral multiplet and they become massive. The product of the two one-loop determinants should then be $1$:
\be
Z_\text{1-loop}^\text{W}(\alpha) = \frac1{Z_\text{1-loop}^{\Phi ,\, q=0}(\alpha)} \;.
\ee
Indeed, $(-1)^{2\delta(\fm)} \prod_\alpha 1/ Z_\text{1-loop}^{\text{chiral},\, q=0} = \prod_\alpha (1-x^\alpha)^{1-g}$. Our regularization consists in introducing a complexified flat connection $v$ for a would-be symmetry that acts on the extra $g$ modes from the vector multiplet as it would on a chiral multiplet. This would be a symmetry of the effective quantum mechanics on $S^1$ obtained by KK reduction on $\Sigma_g$ in each flux sector, if we switched some interactions off. Thus we take as regularized gauge one-loop determinant:
\be
Z_\text{1-loop}^\text{gauge, reg} = (-1)^{2\delta(\fm)} \prod_{\alpha \in G} (1-x^\alpha) \Big( \frac{x^{\alpha/2} y^{1/2} }{ 1-x^\alpha y} \Big)^{-\alpha(\fm) + g} \;.
\ee
Here $\delta = \frac12 \sum_{\alpha>0} \alpha$ is the Weyl vector, and $y=e^{iv}$. For $y \to 1$ we recover $Z_\text{1-loop}^\text{gauge}$, however for $y \neq 1$ the singularities are ``projective'' and we can safely compute the JK residue. At the end of the computation we should take $y \to 1$.

To see what happens, let us consider the simple case of $SU(2)$ Chern-Simons theory at level $k$. We do not need to introduce matter, since we already know how to treat the singularities from chiral multiplets. Applying the rules described above, the classical and regularized one-loop contribution is
\be
Z_\text{cl,1l} = - \frac{(1-x^2)^2}{x^2} \bigg( \frac{ 1-x^2y}{ x^2 - y} \bigg)^{2\fm} \bigg( \frac{ x^2y }{ (1-x^2y)(x^2 - y) } \bigg)^g x^{2k\fm} \;.
\ee
The factor from fermionic zero-modes is
\be
\frac{\partial^2 \log Z_\text{cl,1l} }{ \partial\log x\, \partial \fm } = 2k - \frac{4x^2}{x^2-y} - \frac{4x^2y}{1-x^2y} \;.
\ee
After a redefinition $x^2 = z$, the expression for the partition function is
\be
Z_{\Sigma \times S^1} = - \frac12 \sum_{\fm \in \bZ} \int_\text{JK} \frac{dz}{2\pi i z} \frac{(1-z)^2}z \Big( \frac{ 1-zy}{ z-y} \Big)^{2\fm} \Big( \frac{ zy }{ (1-zy)(z-y) } \Big)^g z^{k\fm} \Big( 2k - \frac{4z}{z-y} - \frac{4zy}{1-zy} \Big)^g
\ee
and the contour is the one prescribed by the JK residue. We choose $\eta<0$, \ie{} we should take minus the residues at $z=0$ and $z=y$ (and not the residues at $z=y^{-1}$ and $z=\infty$). The poles at $z=0$ are confined in the domain $\fm \leq M-1$ (we can take $M$ large and positive), and the poles at $z=y$ in the domain $\fm \geq -M$.
First we compute the sum of the residues at $z=0$, for $\fm \leq M-1$. Let us define
\be
e^{iB} = z^k \Big( \frac{1 - zy}{z - y} \Big)^2 \;.
\ee
The root of the geometric series is $(e^{iB})^{-1}$. Thus, we should find a contour $\cC_a$ that includes only $z=0$ (and not the other poles at $z=y, y^{-1}, \infty$), while being inside the region $|e^{iB}|>1$. On such a contour we have uniform convergence, so we can commute sum and integration. We find
\be
Z_{\Sigma \times S^1}^{(a)} = {\sum_{z_i}}^{(a)} \Res_{z=z_i} \frac{ (1-z)^2}{2z^2} \Big( \frac{zy}{(1-zy)(z-y)} \Big)^g \Big( 2k - \frac{4z}{z-y} - \frac{4zy}{1-zy} \Big)^g \, \frac1{ e^{iB} -1} \;,
\ee
where $z_i$ are the solutions to the BAE
\be
e^{iB} = 1
\ee
inside the contour $\cC_a$. Then we compute the sum of the residues at $z=y$, for $\fm \geq -M$. The root of the geometric series is $e^{iB}$, the inverse than before. This time we should find a contour $\cC_b$ that includes only $z=y$ and lies inside the region $|e^{iB}|<1$. Resumming, we find the same expression as before, but with opposite sign:
\be
Z_{\Sigma \times S^1}^{(b)} = - {\sum_{z_i}}^{(b)} \Res_{z=z_i} \frac{ (1-z)^2}{2z^2} \Big( \frac{zy}{(1-zy)(z-y)} \Big)^g \Big( 2k - \frac{4z}{z-y} - \frac{4zy}{1-zy} \Big)^g \, \frac1{ e^{iB} -1} \;.
\ee
This time $z_i$ are the solutions to the BAE inside the contour $\cC_b$. In practice, solutions that are inside or outside both contours do not contribute---the contour is effectively $\cC_a - \cC_b$.

\begin{figure}[t]
\begin{center}
\includegraphics[width=.5\textwidth]{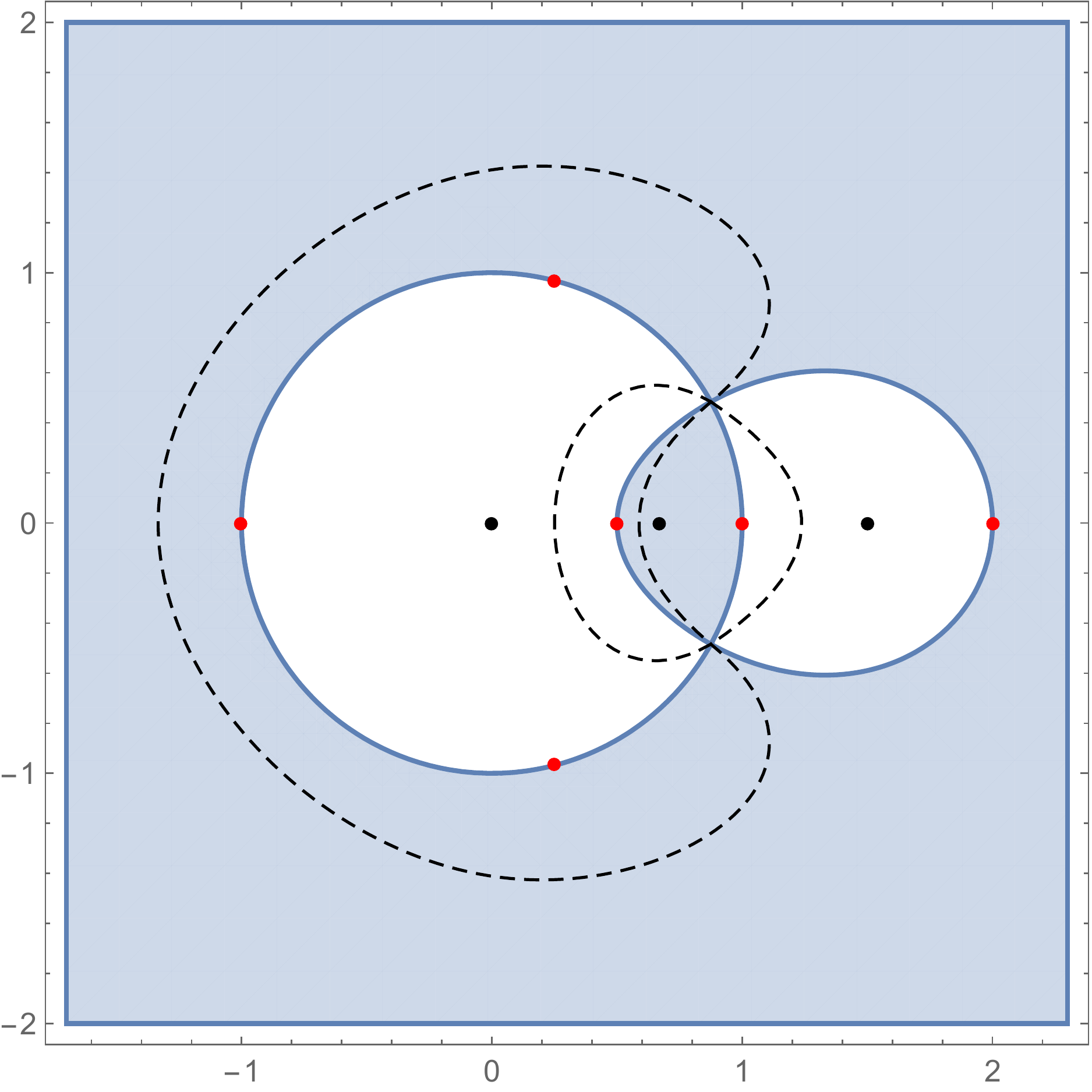}
\caption{Analytic structure of $e^{iB}$ in the complex $z$ plane, for $k=4$ and $y=0.7$. The black dots are the points $\{0, y, y^{-1}\}$ from left to right. The red dots are the solutions to $e^{iB} = 1$. The blue delimiter is the line $|e^{iB}|=1$, while the shaded and white regions have $|e^{iB}| >1$ and $<1$, respectively. The dashed contour inside the shaded region is $\cC_a$, the one inside the white region is $\cC_b$.
\label{fig: CS contours}}
\end{center}
\end{figure}

We should understand the analytic structure of $e^{iB}$. It turns out that we can find the desired contours if we take $y = 1- \epsilon$ with $\epsilon >0$. See Figure \ref{fig: CS contours}. In this case $\cC_a$ contains the $k-1$ approximate roots of $z^k=1$ with $z\neq 1$, as well as a solution along the real interval $(0,y)$; instead $\cC_b$ contains the latter solution and $z=1$. The residue at $z=1$ is zero because of the Vandermonde determinant. Therefore $Z_{\Sigma\times S^1}$ equals the sum of the residues at the $k-1$ approximate roots of $z^k=1$ with $z\neq 1$.

At this point the limit $y\to1$ is taken easily, because all possible subtleties happen at $z=1$ that is excluded, hence the limit is smooth. We find
\be
\label{SU2 regularized sum}
Z_{\Sigma \times S^1} = - {\sum_{z_i}}' \Res_{z=z_i} \, 2k \, \frac{\big[ (1-z)(1-z^{-1}) \big]^{1-g} }{2z} \, \frac1{e^{iB}-1} \;,
\ee
where
\be
e^{iB} = z^k
\ee
and the sum is over the solutions to the BAE $e^{iB}=1$ such that the Vandermonde determinant $\prod_\alpha (1-z^{\alpha/2})$ does not vanish. The expression in (\ref{SU2 regularized sum}) is precisely what one would have obtained from the general formula in (\ref{index formula rank 1}) using the unregularized one-loop determinants, by taking the JK residues from the matter singularities $H_i$ (and not $H_\alpha$), resumming over $\fm$, and then discarding the solutions to the BAEs for which the Vandermonde determinant vanish. Although we have done the computation for $SU(2)$ only, the result is general. It is also clear that the presence of chiral multiplets does not change the argument, as long as their poles are kept away from the zeros of the Vandermonde determinant by a generic choice of flavor background parameters.

\subsection{The final formula}
\label{sec: final formula}

Summarizing, we find the following prescription. One uses (\ref{index formula higher rank})---where only the poles from chiral multiplets and those at infinity are taken into account, as in \cite{Benini:2015noa}---to produce a geometric series in $\fm$ and a contour of uniform convergence. One should resum the geometric series, obtaining a sum of residues at the roots of the BAEs $e^{iB_a} = 1$, but only the roots for which the Vandermonde determinant does not vanish should be kept.

In case all acceptable roots of the BAEs are simple, after some manipulations one obtains, schematically:
\be
\label{index formula schematic -- init}
Z_{\Sigma_g \times S^1} = \frac{(-1)^{\rank G}}{|W|} \; \sum_{x = x_{(i)}} \; Z_\text{cl,1l} \big|_{\fm=0} \; \bigg( \det_{ab} \parfrac{B_a}{u_b} \bigg)^{g-1} \;,
\ee
where
\be
iB_a = \parfrac{ \log Z_\text{cl,1l} }{ \fm_a }
\ee
and $x_{(i)}$ are a set of solutions to the BAEs,
\be
\label{index formula schematic -- fin}
x_{(i)} \qquad\qquad\text{such that}\qquad\qquad e^{iB_a} = 1 \qquad\text{and}\qquad \prod\nolimits_{\alpha\in G} \big(1-x_{(i)}^\alpha \big) \neq 0 \;.
\ee
Notice that in (\ref{index formula schematic -- init}) we could equivalently write the matrix $\partial e^{iB_a} / \partial \log x_b$, since the expression is evaluated at $e^{iB_a}=1$.

This expression has appeared before, for instance in \cite{Okuda:2012nx, Okuda:2013fea, Nekrasov:2014xaa, Gukov:2015sna, Okuda:2015yea}. We have given a precise contour-integral prescription for (\ref{index formula schematic -- init}), which in all cases determines what specific subset of solutions to the BAEs should be kept. Moreover, if the roots of the BAEs are not simple, the more general JK prescription (\ref{index formula higher rank}) should be used.

\subsection{The Witten index}

For $g=1$ and setting the flavor fluxes $\fn=0$, the twisted index reduces to the standard Witten index \cite{Witten:1982df}, \ie{} the supersymmetric partition function of the $\cN=2$ theory on $T^3$. All four supercharges are unbroken, even in the presence of generic twisted masses and flavor flat connections. Hence, it follows from the supersymmetry algebra (\ref{susy algebra}) that the only contributing states are those with $H \pm \sigma^\text{flav} J^\text{flav} = 0$, \ie{} the zero-energy ground states.

As always, we should be careful to interpret the path-integral on $T^3$ as a Witten index (see also the discussion in \cite{Intriligator:2013lca}): this is true only if the theory is well-behaved and it has a finite number of zero-energy states (in particular, no moduli space). We perform our computation with no superpotential, and then treat the superpotential as a $\cQ$-exact deformation that does not affect the path-integral: but this is true only if the theory without superpotential is well-behaved. Thus, we can compute the Witten index of a theory with generic real masses $\sigma^\text{flav}$, provided that those real masses give a finite number of vacua even turning off the superpotential.

This remark is important. Consider the example of a single chiral multiplet $\Phi$ with superpotential $W=\Phi^{N+1}$. This theory has $N$ vacua and the Witten index is $I_\text{W} = N$. The theory has a discrete $\bZ_{N+1}$ flavor symmetry and an R-symmetry with $q_\Phi = 2/(N+1)$. The twisted index from (\ref{Z 1-loop chiral}) is simply
\be
Z = \Big( \frac{ y }{ 1-y} \Big)^{ \frac{1-N}{1+N} (g-1)} \qquad\qquad\text{with}\qquad y^{N+1} = 1 \;.
\ee
For $g=1$ and $y=1$ the formula is not defined, in fact without superpotential the theory has a flat direction and the path integral is not well-defined. For $y \neq 1$ we find, up to a sign, $Z = -1$. This is the correct ``flavored Witten index'' $\Tr\, (-1)^F y^J$, since the $N$ vacua have charges $J = 1, \dots, N$. It is also confirmed by the computation of the elliptic genus in the analogous 2d $\cN=(2,2)$ case \cite{Witten:1993jg}. On the other hand, if we switch off the superpotential the theory has a $U(1)_F$ flavor symmetry, we can turn on the associated real mass $\sigma^\text{flav}$ and then the theory has a single vacuum, $I_\text{W} =1$, in agreement with our formula for $g=1$ and generic $y$.

By the standard argument, the Witten index is independent from continuous deformations of the real masses, and by holomorphy also from the flavor fugacities. Indeed the formula (\ref{index formula schematic -- init})-(\ref{index formula schematic -- fin}) specialized to $g=1$ gives
\be
I_\text{W} = \frac1{|W|}\; \cdot \; \text{\# of acceptable solutions to BAEs} \;,
\ee
the number of acceptable solutions to the BAEs (modded by Weyl transformations).

Another remark is that our computation is reliable only if the theory has a non-anomalous $U(1)_R$ R-symmetry. This is obvious for $g\neq 1$, since we need a background for the R-symmetry to preserve supersymmetry; it is not obvious for $g=1$, \ie{} on flat $T^2 \times S^1$, because there is no background for the R-symmetry and the $U(1)_R$ is not needed to preserve supersymmetry. The point is that in our computation we have assumed that the number of zero-modes equals the one predicted by the index theorem. This is true if we put a generic curved metric on $T^2$, accompanied by a background for the R-symmetry. On the other hand, on flat $T^2 \times S^1$ there are more zero-modes that should be taken into account. This comment is particularly important in the 4d case, where pure SYM breaks the continuous R-symmetry and our formula cannot be applied.


\section{2d theories on \matht{\Sigma_g}}
\label{sec: 2d}

In the two- and four-dimensional cases the analysis is essentially the same, and we will not repeat it here. We will quote the final results.

We consider two-dimensional $\cN=(2,2)$ gauge theories of vector and chiral multiplets with a $U(1)_R$ vector R-symmetry, placed on $\Sigma_g$ with A-twist (same background as in section \ref{sec: 3d}). The Lagrangian is \cite{Witten:1993yc}
\be
\cL = \cL_\text{YM} + \cL_\text{mat} + \cL_{\wt W} + \cL_W \;,
\ee
where the terms are the SYM Lagrangian, the matter kinetic Lagrangian, the twisted superpotential and superpotential interactions, respectively. The parameters of the background are the flux $\fn = \frac1{2\pi} \int_\Sigma F^\text{flav}$ and a constant VEV $v$ for the complex scalar $\sigma^\text{flav}$ in the external vector multiplet. Similarly there are parameters $(\fm,u)$ for the dynamical gauge fields.

We are interested in the partition function of the theory, as well as in correlators of local twisted chiral operators $\cO(x)$ \cite{Witten:1988xj, Witten:1989ig, Witten:1991zz}. These are gauge-invariant polynomial functions $P$ of the complex scalar $\sigma$ in the vector multiplet, which is also the bottom component of a twisted chiral multiplet $\Sigma$. Thus, we are interested is the path-integrals
\be
\big\langle \cO_1(x_1) \dots \cO_s(x_s) \big\rangle_g = \int \cD\varphi \, P_1\big( \sigma(x_1) \big) \dots P_n\big( \sigma(x_n)\big) \, e^{-S}
\ee
on $\Sigma_g$, where $\cO_i = P_i(\Sigma)$. These are usually called ``amplitudes'', they are topological and do not depend on the positions $x_i$.

The classical and one-loop contribution $Z_\text{cl,1l}(u,\fm, v, \fn)$ is constructed as follows. A gauge-invariant twisted superpotential $\wt W(\sigma)$ gives
\be
Z_{\wt W} = e^{4\pi \wt W'(u) \cdot \fm} \;,
\ee
and in the special case that the twisted superpotential is linear and it represents a complexified FI term it becomes
\be
Z_\text{FI} = q^{\Tr \fm}
\ee
where $q = e^{-\zeta - i \theta}$. The one-loop determinant for chiral multiplets is
\be
Z_\text{1-loop}^\text{chiral} = \prod_{\rho \in \fR} \Big[ \frac1{\rho(u)\, \gamma(v)} \Big]^{\rho(\fm) + \gamma(\fn) + (g-1)(q_\rho-1)} \;,
\ee
where $\gamma$ are the flavor weights. For gauge multiplets we have
\be
Z_\text{1-loop}^\text{gauge} = (-1)^{\sum_{\alpha>0} \alpha(\fm)} \prod_{\alpha \in G} \alpha(u) \,\cdot\, du^{\rank G} \;.
\ee
For each operator insertion $\cO_i = P_i(\Sigma)$ we have
$$
P_i(u) \;.
$$
We denote the product of these factors by $Z_\text{cl,1l}$, which is a meromorphic top form on $\fM = \fh_\bC$, the complexified Cartan subalgebra. The contribution from the extra fermionic zero-modes on $\Sigma_g$ is
$$
\bigg( - \det_{ab} \frac{\partial^2 \log Z_\text{cl,1l} }{ \partial u_a \, \partial \fm_b } \bigg)^g \;,
$$
where the sign has been inserted for convenience.

In the Abelian case, the final formula is
\be
Z_{\Sigma_g} = \frac1{|W|} \sum_{\fm \in \Gamma_\fh} \sum_{u_* \in \fM_\text{sing}^*} \JKres_{u = u_*} \big( \sQ_{u_*}, \eta \big) \, \bigg( - \det_{ab} \frac{\partial^2 \log Z_\text{cl,1l} }{ \partial u_a\, \partial \fm_b } \bigg)^g Z_\text{cl,1l} + \text{bound. contrib.}
\ee
The contour at the boundary is controlled by the beta functions for the FI terms. In particular for $G=U(1)$ the boundary contribution is the JK residue at $u=\infty$, assigning charge
\be
Q_\infty = - \sum_i Q_i
\ee
to that point.

In the non-Abelian case the contributions from W-bosons are correctly taken into account by first resumming over $\fm$, and then excluding the roots of the associated BAEs for which the Vandermonde determinant is zero. One obtains a formula as in (\ref{index formula schematic -- init})-(\ref{index formula schematic -- fin}).


\section{4d theories on \matht{\Sigma_g \times T^2}}
\label{sec: 4d}

We consider four-dimensional $\cN=1$ gauge theories of vector and chiral multiplets with a non-anomalous $U(1)_R$ R-symmetry, placed on $\Sigma_g \times T^2$ with twist on $\Sigma_g$. The torus $T^2$ has modulus $\tau$. The Lagrangian is
\be
\cL = \cL_\text{YM} + \cL_\text{mat} + \cL_W \;,
\ee
where the terms are the standard SYM Lagrangian, the matter kinetic Lagrangian, and the superpotential interactions, respectively. For Abelian factors in the gauge group, we can also consider a Fayet-Iliopoulos term
\be
\cL_\text{FI} = - i \frac{\zeta}{2\pi} D \;.
\ee
The parameters of the background are the flux $\fn = \frac1{2\pi} \int_\Sigma F^\text{flav}$ and the flavor flat connection
\be
v = 2\pi \oint_\text{A-cycle} \hspace{-1em} A - 2\pi \tau \oint_\text{B-cycle} \hspace{-1em} A \;.
\ee
Each component of such a variable has identifications $v_a \cong v_a +2\pi \cong v_a + 2\pi \tau$, therefore it lives on a copy of the spacetime $T^2$.
Similarly there are parameters $(\fm, u)$ for the dynamical gauge fields, and $\fM = H^2 \cong T^{2r}$ where $r = \rank G$. It is convenient to define
\be
q = e^{2\pi i \tau} \;,\qquad x = e^{iu} \;,\qquad y = e^{iv}
\ee
with the same notation as before (\ref{Z CS n-Ab}). In particular $x \cong qx$.

The classical and one-loop contribution $Z_\text{cl,1l}$ consists of the following pieces. The only classical-action contribution is from the FI term:
\be
Z_\text{FI} = e^{-\vol(T^2)\, \zeta\, \fm} \;.
\ee
The one-loop determinant for chiral multiplets is
\be
Z_\text{1-loop}^\text{chiral} = \prod_{\rho \in \fR}  \bigg( \frac{i\eta(q)}{\theta_1(x^\rho y^\gamma;q)} \bigg)^{ \rho(\fm) +\gamma(\fn) +(g-1) (R_\rho-1)} \;,
\ee
where the  elliptic functions are $\eta(q) = q^{1/24} \prod_{n=1}^\infty (1-q^n)$ and
\be
\theta_1(x;q) = -i q^\frac18 x^\frac12 \prod_{k=1}^\infty (1-q^k)(1-xq^k)(1-x^{-1}q^{k-1}) = -i \sum_{n\in\bZ} (-1)^n e^{iu( n + \frac12)} \, e^{\pi i \tau (n+\frac12)^2} \;.
\ee
We used $R$ for the R-charges, not to make confusion with the modular parameter $q$.
The one-loop determinant for off-diagonal vector multiplets is
\bea
 Z_\text{1-loop}^\text{gauge, off} =  (-1)^{\sum_{\alpha>0} \alpha(\fm)} \, \bigg( \, \prod_{\alpha\in G} \frac{\theta_1\big(x^\alpha ;q \big) }{i\eta(q)}\bigg)^{1-g} \;,
\eea
while the contribution from vector multiplets along the Cartan generators is
\be
Z_\text{1-loop}^\text{gauge, Cartan} = \eta(q)^{2r(1-g)} \, (i\, du)^r \;. 
\ee
The contribution from the fermionic zero modes on $\Sigma_g$ is
$$
\bigg ( \det_{ab} \frac{\partial^2 \log Z_\text{1loop}}{\partial iu_a\,\partial \fm_b}\bigg )^g \;.
$$

In the Abelian case the final formula is
\be
Z_{\Sigma_g \times T^2} = \frac1{|W|} \sum_{\fm \in \Gamma_\fh} \sum_{u_* \in \fM_\text{sing}^*} \JKres_{u = u_*} \big( \sQ_{u_*}, \eta \big) \, \bigg( \det_{ab} \frac{\partial^2 \log Z_\text{cl,1l} }{ \partial iu_a\, \partial \fm_b } \bigg)^g Z_\text{cl,1l} \;.
\ee
In particular there are no boundary contributions since the integration domain $\fM$ is compact.

In the non-Abelian case the contributions from W-bosons are correctly taken into account by first resumming over $\fm$, and then excluding the roots of the associated BAEs for which the Vandermonde determinant is zero.%
\footnote{In the four-dimensional case there is a small subtlety: the $U(1)$ symmetry associated to an adjoint chiral multiplet, used in the regularization argument we gave in section \ref{sec: derivation}, is anomalous. We can neglect this problem because we only need to turn on an infinitesimal deformation, and then turn it off at the end of the computation.}
One obtains a formula as in (\ref{index formula schematic -- init})-(\ref{index formula schematic -- fin}).


\section{Examples}
\label{sec: examples}

In this section we present various examples, to illustrate the use of the formula. We perform new checks of non-perturbative dualities, and compare with known results when available. We will compactly call $Z$ the partition function $Z_{\Sigma_g \times S^1}$.

\subsection[$U(1)_k$ Chern-Simons theory]{\matht{U(1)_k} Chern-Simons theory}

We start with $U(1)$ supersymmetric Yang-Mills-Chern-Simons (YMCS) theory at level $k$. At low energies this is equivalent to bosonic Chern-Simons at level $k$. The classical and one-loop contribution is $Z_\text{cl,1l} = x^\ft \xi^\fm x^{k\fm}$, turning on a background for the topological symmetry. Thus
\be
Z = \sum_{\fm \in \bZ} \int_\text{JK} \frac{dx}{2\pi i x} \, k^g x^{k\fm + \ft} \xi^\fm \;.
\ee
The charges of the points at infinity are $Q_0 = -k$ and $Q_\infty = k$. Assuming $k>0$, we can choose $\eta<0$ and pick minus the residue at $x=0$. This gives
\be
Z = \begin{cases} - k^g \, \xi^{-\ft/k} \qquad &\text{if } \ft=0 \pmod{k} \;,\\ 0 &\text{otherwise} \;. \end{cases}
\ee
For $\ft=0$ and up to an ambiguous sign, this gives the known result $k^g$, which is the number of ground states of $U(1)_k$ Chern-Simons on $\Sigma_g$.

\subsection[$U(1)_{1/2}$ with one chiral multiplet]{\matht{U(1)_{1/2}} with one chiral multiplet}

We consider the following duality \cite{Dimofte:2011ju, Benini:2011mf}. The ``electric theory'' is a $U(1)$ YMCS theory at level $k=\frac12$ with one chiral multiplet of gauge charge $1$ and R-charge $1$. The ``magnetic theory'' is a free chiral multiplet with flavor charge $1$ under $U(1)_T$ and R-charge $0$, and global CS terms $k_{TT} = - \frac12$, $k_{RT} = - \frac12$.

In the electric theory, the classical and one-loop contribution is
\be
Z_\text{cl,1l} = x^\ft (-\xi)^\fm x^{\fm/2} \Big( \frac{ x^{1/2} }{ 1-x} \Big)^\fm \;,
\ee
where we have redefined the sign of $\xi$ for later convenience. Thus, the index is given by
\be
Z = \sum_{\fm\in\bZ} \int_\text{JK} \frac{dx}{2\pi i x} \, \frac{(-\xi)^\fm x^{\ft + \fm} }{ (1-x)^{\fm + g}} \;.
\ee
There are poles at $x=0, 1, \infty$, with associated charges $-1, 1, 0$, respectively. We can choose $\eta = -1$, therefore we pick minus the residues at $x=0$. We resum $\sum_{\fm \leq M-1}$ for some large $M$. If we take a contour around $x=0$ and $|\xi|\gg1$, we have uniform convergence on the contour and we can exchange summation and integration. The resummed function has a unique pole inside the contour, at $x= (1-\xi)^{-1}$. Hence:
\be
Z = - \oint_{x = \frac1{1-\xi}} \frac{dx}{2\pi i} \, \frac{x^{\ft -1}}{(1-x)^{g-1}} \, \frac{ \big( \frac{\xi x}{x-1} \big)^M }{ (1-\xi)x - 1} = (-1)^g \, \frac{\xi^{1-g} }{ (1-\xi)^{\ft + 1 - g}} \;.
\ee
Up to the ambiguous sign, this is precisely the index of the magnetic theory:
\be
Z = (-1)^g \, \Big( \frac{\xi^{1/2}}{1-\xi} \Big)^{\ft + 1 - g} \xi^{-\ft/2} \xi^{- (g-1)/2} \;.
\ee

\subsection[Witten index of $U(1)_k$ with matter]{Witten index of \matht{U(1)_k} with matter}

Consider $U(1)_k$ with $N$ chiral multiplets of charges $Q_i$ and R-charge $1$ (in order to avoid parity anomalies in the R-symmetry).
The classical and one-loop contribution is
\be
Z_\text{cl,1l} = \xi^\fm x^{k\fm + \ft} \prod_{i=1}^N \bigg( \frac{x^{Q_i/2} y_i^{1/2} }{ 1-x^{Q_i} y_i} \bigg)^{Q_i \fm + \fn_i} \;,
\ee
where $(y_i, \fn_i)$ control the background for flavor symmetries up to one combination that could be reabsorbed into $(x,\fm)$, and $(\xi,\ft)$ controls the background for the topological symmetry. In order to avoid a gauge-gauge parity anomaly (\ie{} in order for $Z_\text{cl,1l}$ to be a single-valued function of $x$) the condition
\be
k + \frac12 \sum\nolimits_i Q_i^2 \in \bZ \qquad\text{equivalent to}\qquad k + \frac12 \sum\nolimits_i Q_i \in \bZ
\ee
should be met. In general there are gauge-flavor parity anomalies: one could cure them with the addition of gauge-flavor CS terms, however to the purpose of computing the Witten index we will simply set $\fn_i= \ft=0$. In general there are also flavor-flavor parity anomalies, however they are not inconsistencies of the theory: they simply imply that $Z$ is not a single-valued function of $y_i$.

The poles are at $x=0, y_i^{-1/Q_i}, \infty$. Depending on the choice of $\eta = \pm1$ one will have to collect different classes of residues. In any case, the sum over $\fm$ generates the expression in (\ref{index formula schematic -- init}):
\be
\label{index U(1)k N fund}
Z = - \sum_{x = x_{(\alpha)}} Z_\text{cl,1l} \big|_{\fm=0} \, \bigg( \parfrac{e^{iB}}{\log x} \bigg)^{g-1} \;,
\ee
where $x_{(\alpha)}$ are the roots of the BAE
\be
e^{iB} = \xi \, x^k \prod_{i=1}^N \bigg( \frac{x^{Q_i/2} y_i^{1/2} }{ 1-x^{Q_i} y_i} \bigg)^{Q_i} = 1 \;.
\ee
If we specialize to $g=1$ and $\fn_i=\ft =0$, the index $Z_{T^3}$ is the Witten index $I_\text{W}$ of the theory. On the other hand (\ref{index U(1)k N fund}) becomes, up to the sign, the number of solutions to the BAE.

To compute the number of solutions, we divide the chiral multiplets into two groups: $I_+$ are those with $Q_i>0$, $I_-$ are those with $Q_i <0$. Then the equation can be written as
\be
\label{equation 1}
\Big( \xi \prod\nolimits_i y_i^{Q_i/2} \Big) \;\; x^{k + \frac12 \sum_{i\in I_+} Q_i^2 - \frac12 \sum_{i\in I_-} Q_i^2} \; \prod_{i \in I_-} \big( x^{|Q_i|} - y_i \big)^{|Q_i|} = \prod_{i\in I_+} \big( 1 - x^{Q_i} y_i \big)^{Q_i} \;.
\ee
We define the non-negative numbers
\be
n_+ = \frac12 \sum_{i\in I_+} Q_i^2 \;,\qquad\qquad n_- = \frac12 \sum_{i \in I_-} Q_i^2 \;.
\ee
Then the number of solutions to (\ref{equation 1}) is
\be
I_\text{W} = \begin{cases} \max\big( k + n_+ + n_- \,,\, 2n_+ \big) \qquad &\text{if } k + n_+ - n_- \geq 0 \;, \\ \max\big( -k + n_+ + n_- \,,\, 2n_- \big) \qquad &\text{if } k + n_+ - n_- \leq 0 \;. \end{cases}
\ee
After some algebra, this expression can be rewritten as
\be
I_\text{W} = \begin{cases} |k| + n_+ + n_- \qquad &\text{if } |k| \geq |n_+ - n_-| \;, \\ \max (2n_+ \,,\, 2n_-) \qquad &\text{if } |k| \leq |n_+ - n_-| \;. \end{cases}
\ee
This reproduces the Witten index computed in \cite{Intriligator:2013lca}.

\subsection[$U(N_c)$ SQCD with $N_f$ flavors and Aharony duality]{\matht{U(N_c)} SQCD with \matht{N_f} flavors and Aharony duality}

Consider a $U(N_c)$ theory, with $N_f$ chiral multiplets $Q_a$ in the fundamental and $\tilde Q_b$ in the antifundamental representations, and no CS interactions. The topologically twisted index of this theory for $g=0$ has been computed in \cite{Benini:2015noa}, and matched with the index of its Aharony dual theory \cite{Aharony:1997gp}. Here we generalize that computation to arbitrary $g$.

For simplicity, we only introduce backgrounds for the R-symmetry, the topological symmetry and the axial $U(1)_A$ subgroup of the flavor symmetry acting with the same charge on all chiral fields. We use $(\xi, \ft)$ for the fugacity and background flux for the topological symmetry, and $(y,\fn)$ for the axial flavor symmetry. We assign R-charge $1$ to the chiral fields. Hence:
\be\nn
\begin{array}{c|cccc}
 & U(N_c) & U(1)_T & U(1)_A & U(1)_R \\
\hline
Q_a & \rep{N_c} & 0 & 1 & 1 \\
\tilde Q_b & \rep{\overline N_c} & 0 & 1 & 1 \\
\hline
M_{ab} = Q_a \tilde Q_b & \rep{1} & 0 & 2 & 2 \rule{0pt}{1.1em} \\
T & \rep{1} & 1 & -N_f & -N_c+  1 \\
\tilde T & \rep{1} & -1 & -N_f & -N_c + 1
\end{array}
\ee
Here $T, \tilde T$ are the monopole operators $V_\fm$ corresponding to magnetic fluxes $\fm = (1,0, \dots ,0)$ and $\fm = (0, \dots 0, -1)$, respectively.

The classical and one-loop contribution is
\be
Z_\text{cl,1l} = (-1)^{N_c} \prod_{a=1}^{N_c} x_a^\ft \big( (-1)^{N_f} \xi \big)^{\fm_a} \bigg( \frac{x_a^{1/2} y^{1/2} }{ 1-x_a y} \bigg)^{N_f(\fm_a + \fn)} \bigg( \frac{ x_a^{-1/2} y^{1/2} }{ 1-y/x_a} \bigg)^{N_f(\fn-\fm_a)} \, \prod_{a\neq b}^{N_c} \Big( 1 - \frac{x_a}{x_b} \Big)^{1-g} \;.
\ee
We have redefined $\xi \to (-1)^{N_f}\xi$ and included a sign $(-1)^{N_c}$ for later convenience. We directly apply the formula in (\ref{index formula schematic -- init})-(\ref{index formula schematic -- fin}). The quantities $B_a$ are given by
\be
iB_a = \parfrac{\log Z_\text{cl,1l}}{\fm_a} = \log \, (-1)^{N_f}\xi + N_f \log \bigg( \frac{x_a^{1/2} y^{1/2} }{ 1-x_a y} \bigg) - N_f \log \bigg( \frac{ x_a^{-1/2} y^{1/2} }{ 1-y/x_a} \bigg) \;.
\ee
The BAEs are
\be
1 = e^{iB_a} = \xi \bigg( \frac{y-x_a}{1-x_a y} \bigg)^{N_f} \;.
\ee
Since the BAEs are all equal and decoupled, the solutions are simply collections of roots of the polynomial
\be
\cP(x) = \xi (y - x)^{N_f} - (1-xy)^{N_f} \equiv (-1)^{N_f} (\xi - y^{N_f}) \prod_{\alpha=1}^{N_f} (x- x_\alpha) \;,
\ee
where in the last expression we have written $\cP$ in terms of its $N_f$ roots $x_\alpha$. According to the prescription, one should only retain solutions in which the $N_c$ selected roots $\{x_\alpha\}$ are all different (otherwise the Vandermonde determinant vanishes). This leads to a sum over ordered collections of $N_c$ distinct roots out of $N_f$. The contribution from the fermionic zero-modes on $\Sigma_g$ follows from the matrix
\be
\parfrac{iB_a}{\log x_b} = \delta_{ab} N_f \, \frac{1-y^2}{(1-x_ay)(1-y/x_a)} \;.
\ee
Putting all pieces together, we find the expression
\be
\label{U(N_c) first expression}
Z = \frac{ y^{N_c N_f \fn} N_f^{N_c(g-1)} }{ (1-y^2)^{N_c (1-g)} } \; \sum_I \; \prod_{\alpha \in I} \frac{ x_\alpha^{\ft + (N_c-1)(g-1)} }{ (1-x_\alpha y)^{N_f \fn + g-1} (1-y/x_\alpha)^{N_f \fn + g-1} } \prod_{\substack{\beta \in I \\ (\beta \neq \alpha)}} (x_\alpha - x_\beta)^{1-g} \;.
\ee
The sum over $I$ runs over the set $C^{N_f}_{N_c}$ of unordered combinations of $N_c$ different integers in $\{ 1,\ldots,N_f\}$, while $I^c$ denotes the complementary set  $\{ 1,\cdots,N_f\}\setminus I$ belonging to $C^{N_f}_{N_f-N_c}$. We used that the summands in (\ref{U(N_c) first expression}) are invariant under permutations of the roots, to reduce the sum to unordered combinations and cancel the factor $N_c!$.

The number of combinations is
\be
I_\text{W} = \#\, C^{N_f}_{N_c} = \binom{N_f}{N_c} \qquad\quad\text{if} \quad N_f \geq N_c
\ee
and zero otherwise, which gives the Witten index of the theory (when well-defined), setting $g=1$ and $\fn = \ft = 0$. In fact the theory breaks supersymmetry for $N_f \leq N_c-2$, is IR free (therefore the Witten index is not well-defined) for $N_f = N_c-1$, and has isolated vacua for $N_f \geq N_c$.

To make contact with \cite{Benini:2015noa}, we notice the following identities:
\be
\label{full products A}
\prod_{\alpha=1}^{N_f} x_\alpha = \frac{\xi y^{N_f} -1}{ \xi - y^{N_f}} \;,\qquad\qquad \prod_{\alpha=1}^{N_f} (1-x_\alpha y) = \frac{\xi (1-y^2)^{N_f}}{ \xi - y^{N_f}} \;,
\ee
and
\be
\prod_{\beta (\neq \alpha)}^{N_f} (x_\alpha - x_\beta) = \frac{N_f \xi (1-y^2) }{ \xi - y^{N_f}} \;\; \frac{ x_\alpha^{N_f-1} (1-y/x_\alpha)^{N_f-1} }{ 1-x_\alpha y} \;.
\ee
They can be used to recast the index into the form
\be
\label{partitionNcNf}
Z = \frac{(-1)^{N_c N_f (\fn+g-1)} \xi^{N_c \fn} y^{N_c N_f \fn} }{ (\xi - y^{N_f})^{N_c(1-g)} } \; \sum_I \; \prod_{\alpha \in I} \frac{ x_\alpha^{\ft + N_c(g-1) + N_f\fn} }{ (1-x_\alpha y)^{N_f(2\fn+g-1)} \prod_{\beta \in I^c} (x_\alpha - x_\beta)^{1-g} } \;.
\ee

For $N_f < N_c$, the expression above obviously vanishes.
If $N_f = N_c$, there is only one $I$ while $I^c = \emptyset$. We immediately get
\be
\label{dualNc}
Z_{N_f = N_c} = (-1)^\ft \; \frac{ y^{N_c^2(3\fn +2g - 2)} \; \xi^\ft }{ (1-y^2)^{N_c^2 (2\fn + g - 1)} \; (1-\xi y^{-N_c})^{N_c(1 - g -\fn) + \ft} \; (1-\xi^{-1} y^{-N_c})^{N_c(1 - g -\fn)-\ft} } \;.
\ee
The dual theory for $N_f=N_c$ is given by the fields $M_{ab}$, $T$ and $\tilde T$, coupled through the superpotential $W = T\tilde T \det M$ \cite{Aharony:1997bx}. The partition function of the dual theory is then
\be
Z_\text{dual}^{N_f = N_c} = \Big( \frac y{1-y^2} \Big)^{(2\fn + g -1)N_c^2} \Big( \frac{\xi^\frac12 y^{-\frac{N_c}2} }{1-\xi y^{-N_c}} \Big)^{N_c(1 - g -\fn) + \ft} \Big( \frac{ \xi^{-\frac12} y^{-\frac{N_c}2} }{ 1-\xi^{-1} y^{-N_c} } \Big)^{N_c(1 - g -\fn) - \ft} \;.
\ee
This agrees with (\ref{dualNc}), up to an ambiguous sign $(-1)^\ft$.

The expression (\ref{partitionNcNf}) for $N_f>N_c$ is more complicated but we can use it  to check Aharony dualities \cite{Aharony:1997gp}.
The dual theory is a $U(N_f-N_c)$ gauge theory with $N_f$ fundamentals $q_a$, $N_f$ anti-fundamentals $\tilde q_b$ and $N_f^2+2$ singlets $M_{ab}$,  $T$ and $\tilde T$, corresponding to the mesons and monopoles of the original theory, with a superpotential $W = M_{ab} q_a \tilde q_b + v_- T + v_+ \tilde T$, where $v_{\pm}$ are monopoles of the dual theory. We assign the charges consistently with the original theory:
\be\nn
\begin{array}{c|cccc}
 & U(N_f-N_c)_g & U(1)_T & U(1)_A & U(1)_R \\
\hline
q_a & \rep{N_f-N_c} & 0 & -1 & 0 \\
\tilde q_b & \overline{\rep{N_f - N_c}} & 0 & -1 & 0 \\
M_{ab} & 0 & 0 & 2 & 2 \\
T & 0 & 1 & -N_f & -N_c+1 \\
\tilde T & 0 & -1 & -N_f & -N_c+1\\
\hline
v_+ & 0 & 1 & N_f & N_c+1 \\
 v_-& 0 & -1 & N_f & N_c+1
\end{array}
\ee
Notice that the dual quarks have R-charge zero and axial flavor charge $-1$.

The partition function of the dual theory is obtained by multiplying the contribution of the gauge sector for the quarks $q_a,\tilde q_b$ with the contribution of the singlets  $M_{ab}$, $T$ and $\tilde T$. The first contribution is the partition function for a  $U(N_f-N_c)$ theory with quarks $q_a,\tilde q_b$ which we can read from (\ref{partitionNcNf}). According to our assignment of charges, we need to replace the background charge and fugacity for the flavor symmetry  by $y \leftrightarrow y^{-1}$ and $\fn \leftrightarrow 1 - g -\fn$, as well as $N_c \leftrightarrow N_f - N_c$. We find
\be
\label{Z q qtilde}
Z_{q\tilde q} = \frac{y^{(N_f - N_c)N_f(\fn + g -1)} \xi^{(N_f - N_c)(1 - g -\fn)} }{ (-1)^{(N_f - N_c)N_f\fn} (\xi - y^{-N_f} )^{(N_f - N_c)(1-g)} } \; \sum_J \; \prod_{\beta \in J} \frac{\tilde x_\beta^{-N_f \fn + N_c(1-g) + \ft} }{ (1-\tilde x_\beta y^{-1})^{N_f (1-g-2\fn)} \prod\limits_{\alpha \in J^c} (\tilde x_\beta - \tilde x_\alpha)^{1-g} }
\ee
and the sum is over $J \in C^{N_f}_{N_f-N_c}$.
The $\tilde x_\beta$ are the roots of $\tilde\cP(\tilde x) = \xi(y^{-1} - \tilde x)^{N_f} - (1-\tilde x y^{-1})^{N_f} = 0$, and in fact $\tilde x_\beta = 1/x_\beta$. We can thus rewrite $Z_{q\tilde q}$ in terms of $x_\beta$, and convert the products over $J$ into products over $J^c$ using the full products in (\ref{full products A}). We get:
\begin{multline}
Z_{q\tilde q} = \frac{(-1)^{N_f(N_f-N_c)(1-g-\fn)} y^{N_f(N_f-N_c)(1-g-\fn)} \xi^{N_f\fn + N_c(\fn + g -1)} }{ (1-y^2)^{N_f^2(1-g-2\fn)} (\xi-y^{N_f})^{N_f\fn - \ft} (\xi y^{N_f}-1)^{N_f\fn - N_c(1-g) + \ft} } \\
\times \sum_J \prod_{\alpha \in J^c} \frac{ x_\alpha^{N_f\fn - N_c(1-g) + \ft} }{ (1-x_\alpha y)^{N_f(2\fn + g -1)} \prod_{\beta\in J} (x_\alpha - x_\beta)^{1-g} } \;.
\end{multline}
The contribution of the gauge singlets is
\bea
Z_{MT\tilde T} &= \Big( \frac y{1-y^2} \Big)^{N_f^2(2\fn + g -1)} \Big( \frac{\xi^\frac12 y^{-\frac{N_f}2} }{ 1-\xi y^{-N_f} } \Big)^{\ft - N_f \fn+N_c(1-g) } \Big( \frac{\xi^{-\frac12} y^{-\frac{N_f}2} }{1-\xi^{-1} y^{-N_f} } \Big)^{-\ft - N_f \fn + N_c(1-g) } \\
&= \frac{ (-1)^{\ft - N_f\fn + N_c(1-g) } y^{N_f \left( N_f(\fn + g -1) + N_c(1-g) \right)} \xi^{-N_f\fn + N_c(1-g)} }{ (1-y^2)^{N_f^2(2\fn + g -1)} (\xi - y^{N_f})^{\ft - N_f\fn + N_c(1-g) } (\xi y^{N_f} -1)^{-\ft - N_f\fn + N_c(1-g) } } \;.
\eea
Then the partition function of the dual theory, $Z_\text{dual} = Z_{q\tilde q} Z_{MT\tilde T}$, equals the one of the electric theory up to $(-1)^{(N_f - N_c)(g-1) + \ft}$.

Since Giveon-Kutasov duality \cite{Giveon:2008zn} can be derived from Aharony duality with an RG flow \cite{Giveon:2008zn, Benini:2011mf} (and viceversa \cite{Intriligator:2013lca}), we have also implicitly verified that Giveon-Kutasov dual theories have the same higher-genus index.

\subsection[$SU(2)_k$ Chern-Simons theory]{\matht{SU(2)_k} Chern-Simons theory}

$SU(2)$ supersymmetric YMCS theory at level $k \geq 3$ is equivalent, at low energies, to bosonic $SU(2)$ CS at level $\bar k = k-2$ (while for $k=0,1$ it breaks supersymmetry, and for $k=2$ it confines; we assume $k\geq 0$).

The classical and one-loop contribution is
\be
Z_\text{cl,1l} = x^{2k\fm} \big[ (1-x^2)(1-x^{-2}) \big]^{1-g} \;.
\ee
The naive expression for the partition function is
\be
Z = \frac{(-1)^{g-1}}2 \sum_{\fm \in \bZ} \int_\text{JK} \frac{dx}{2\pi i x} \, (2k)^g x^{2k\fm} \bigg[ \frac{(1-x^2)^2}{x^2} \bigg]^{1-g} \;,
\ee
however this expression does not correctly capture the contribution from W-bosons. The effective charges at the boundary are $Q_0 = -k$ and $Q_\infty = k$. We choose $\eta<0$, therefore we should pick minus the residues at $x=0$. A contour of uniform convergence for the geometric series in $\fm$ is in $|x|>1$. According to the prescription in section \ref{sec: final formula}, we should first resum over $\fm \leq M-1$ for some large $M$, and then pick the residues inside the contour for which the Vandermonde determinant does not vanish. The geometric series has poles at the roots of $x^{2k} = 1$. Thus, after some manipulations, we find
\be
Z = (-1)^g \frac{(2k)^{g-1}}2 \sum_{x=x_{(i)}} \bigg[ \frac{(1-x^2)^2}{x^2} \bigg]^{1-g} \;,
\ee
where $x_{(i)}$ are the roots of $x^{2k}=1$ with $x_{(i)}^2 \neq 1$. This is precisely the expression in (\ref{index formula schematic -- init})-(\ref{index formula schematic -- fin}), using $B = 2k u$.

The index can be recast, up to the ambiguous sign, in the form
\be
Z = \Big( \frac{\bar k + 2}2 \Big)^{g-1} \sum_{j=1}^{\bar k + 1} \Big( \sin \frac{\pi i}{\bar k+2} j \Big)^{2-2g} \;,
\ee
which is the standard Verlinde formula for $SU(2)_{\bar k}$.

\subsection[$SU(2)_k$ with matter and the ``duality appetizer'']{\matht{SU(2)_k} with matter and the ``duality appetizer''}

Let us consider $SU(2)_k$ with matter. First, we take one chiral multiplet $\Phi$ in the adjoint representation (this setup has been studied at length in \cite{Gukov:2015sna}). There is a flavor $U(1)_F$ symmetry that rotates the adjoint, and we indicate by $(y,\fn)$ the corresponding background. To cancel a parity anomaly we introduce a flavor-flavor CS term $k_{FF} = \frac12$. We assign R-charge $1$ to $\Phi$. The classical and one-loop contribution is
\be
Z_\text{cl,1l} = - \frac{(1-x^2)^{2-2g}}{x^{2-2g}} \, x^{2k\fm} \, y^{\fn/2} \, \bigg( \frac{xy^{1/2}}{1-x^2y} \bigg)^{2\fm+\fn} \bigg( \frac{y^{1/2}}{1-y} \bigg)^\fn \bigg( \frac{x^{-1} y^{1/2} }{ 1-x^{-2}y} \bigg)^{-2\fm + \fn} \;,
\ee
where we included a sign $(-1)^g$ for later convenience. The associated BAE is
\be
e^{iB} = x^{2k} \bigg( \frac{x^2 - y}{1-x^2 y} \bigg)^2 = 1 \;.
\ee
This equation has $2|k| + 4$ solutions, however two of them are $\pm1$ and should be discarded. The index takes the general form (\ref{index formula schematic -- init})-(\ref{index formula schematic -- fin}).

For $k=1$ we can evaluate the index explicitly. The four acceptable solutions to the BAE are $x^2 = \frac12 \big( y^2 + 2y - 1 \pm (y+1) \sqrt{ y^2 + 2y - 3} \big)$, and substituting we obtain
\be
Z = 2^g \, y^{2\fn} \, (y^2 -1)^{1-g-2\fn} \;.
\ee
In fact, it has been argued in \cite{Jafferis:2011ns, Intriligator:2013lca} that the theory is dual to a free chiral multiplet $Y = \Tr \Phi^2$ with flavor charge $2$ and R-charge $2$, plus an R-flavor CS term $k_{RF} = -1$ and a topological sector $U(1)_2$. The index of the dual theory is
\be
Z_\text{dual} = 2^g \, y^{1-g} \, \Big( \frac y{1-y^2} \Big)^{2\fn + g -1} \;,
\ee
which is the same as before up to an ambiguous sign. Notice how the higher-genus index captures, through the factor $2^g$, the topological sector.

We can consider other types of matter content. For instance, let us take $2N_f$ chiral multiplets in the fundamental representation (to cancel Witten's anomaly \cite{Witten:1982fp} the number should be even). Introducing for simplicity a background for the axial $U(1)_A$ flavor symmetry only, and assigning to the flavors R-charge $1$, the classical and one-loop contribution is
\be
Z_\text{cl,1l} = x^{2k\fm} \big[ (1-x^2)(1-x^{-2}) \big]^{1-g} \bigg( \frac{x^\frac12 y^\frac12}{1-xy} \bigg)^{2N_f(\fm+\fn)} \bigg( \frac{x^{-\frac12} y^\frac12}{1-x^{-1}y} \bigg)^{2N_f(-\fm+\fn)} \;.
\ee
The associated BAE is
\be
e^{iB} = x^{2k} \Big( \frac{x-y}{1-xy} \Big)^{2N_f} =1 \;.
\ee
This equation has $2|k| + 2N_f$ solutions, however two of them are $\pm1$ which should be discarded because are zeros of the Vandermonde determinant. We thus have $2|k| + 2N_f -2$ acceptable solutions $x_{(i)}$. Expanding out (\ref{index formula schematic -- init})-(\ref{index formula schematic -- fin}) we get
\be
Z = \frac{(-1)^g}2 \sum_{x = x_{(i)}} \frac{(1-x^2)^{2-2g}}{x^{2-2g}} \, \frac{y^{2N_f\fn} x^{2(N_f + k)\fn} }{ (1-xy)^{4N_f \fn} } \bigg( 2k + \frac{2N_f (1-y^2) }{ (1-xy)(1-x^{-1}y)} \bigg)^{g-1} \;.
\ee
Specializing to $g=1$ and $\fn=0$, we obtain the Witten index of the theory equal to half the number of acceptable solutions to the BAE:
\be
I_\text{W} = |k| + N_f - 1
\ee
or $I_\text{W} = 0$ if the number on the right-hand-side is negative, in agreement with \cite{Intriligator:2013lca}.

In case of arbitrary matter content $\fR = \bigoplus \fR_i$, where $\fR_i$ is the spin $I_i$ representation, it is easy to repeat the computation of the Witten index. Each representation $\fR_i$ brings a factor
$$
\prod_{j=0}^{\big\lfloor I_i - \frac12 \big\rfloor} \bigg( \frac{x^{2(I_i - j)} - y }{ 1 - x^{2(I_i-j)} y } \bigg)^{2(I_i - j)}
$$
to the BAE. Thus half the number of acceptable solutions is
\be
I_\text{W} = |k| + \frac12 \sum_i T_2(\fR_i) - 1 \;,
\ee
where $T_2$ is the quadratic Casimir $T_2(\fR_i) = \frac23 (2I_i +1) I_i (I_i+1)$. This agrees with \cite{Intriligator:2013lca}.

\subsection{Two-dimensional A-twisted $\bC\bP^{N-1}$ model}

We consider a two-dimensional $\cN{=}(2,2)$ $U(1)$ theory with $N$ chiral multiplets with charge $1$ (and R-charge $0$). In the IR this theory realizes a NLSM on $\bC\bP^{N-1}$. We want to compute correlators of the field-strength twisted chiral multiplet $\Sigma$, which represents the K\"ahler class (also known as hyperplane class) of $\bC\bP^{N-1}$. With $s$ insertions of $\Sigma$, the classical and one-loop contribution is
\be
Z_\text{cl,1l} = u^s \, q^\fm \, \frac1{u^{(\fm + 1 - g)N}} \;.
\ee
Therefore we find the expression
\be
\Big\langle \Sigma(x_1) \dots \Sigma(x_s) \Big\rangle_g = \sum_{\fm \in \bZ} \int_\text{JK} \frac{du}{2\pi i} \, \frac{N^g \, q^\fm}{ u^{(\fm +1-g)N + g - s} } \;.
\ee
We can choose $\eta >0$, then we should take the residues of the poles at $u=0$ from chiral multiplets. The result is
\be
\big\langle \Sigma_1 \dots \Sigma_s \big\rangle_g = \begin{cases} N^g \, q^{\frac{ s + (N-1)(g-1)}N} \qquad &\text{if } s = g-1 \pmod{N} \;, \\ 0 &\text{otherwise} \;, \end{cases}
\ee
where we indicated $\Sigma(x_p)$ simply as $\Sigma_p$ since the amplitudes are independent from the positions.

For $g=0$ the shortest non-vanishing correlator is $\langle \Sigma_1 \dots \Sigma_{N-1} \rangle_{g=0} = 1$, corresponding to the fact that the intersection of $N-1$ hyperplanes in $\bC\bP^{N-1}$ is a single point, while higher-point correlators are determined by the quantum cohomology (or chiral ring) relation $\Sigma^N = q$. For $g=1$ the shortest non-vanishing correlator is
\be
\langle 1 \rangle_{g=1} = N \;,
\ee
which reproduces the Witten index of $\bC\bP^{N-1}$.

All other correlators follow from the fact that we have a topological field theory \cite{Witten:1989ig}. The $N$ states of the theory on $S^1$ are realized by insertions of $\Sigma^k$ with $k=0, \dots, N-1$, and their duals by $\Sigma^{N-k-1}$. Indeed the correlator $\langle \Sigma^j \, \Sigma^{N-k-1} \rangle_{g=0} = \delta_{jk}$ (with $j,k=0, \dots, N-1$) is interpreted as the propagator. Higher genus correlators are obtained by adding two insertions of $\Sigma^k$, $\Sigma^{N-k-1}$ and gluing with a propagator. The unpunctured torus is $\langle 1 \rangle_{g=1} = \sum_k \langle \Sigma^k \, \Sigma^{N-k-1} \rangle_{g=0} = N$. The case of $g=2$ with one puncture is $\langle \Sigma \rangle_{g=2} = \sum_k \langle \Sigma\, \Sigma^k \, \Sigma^{N-k-1} \rangle_{g=1} = N^2 q$, where we used the chiral ring relation. All other correlators can be reproduced this way.

This example could be generalized in many ways. For instance one could study a $U(N_c)$ gauge theory with $N_f$ chiral multiplets in the fundamental representation---which flows to a NLSM on the complex Grassmannian $Gr(N_c,N_f)$---and possibly $N_a$ in the antifundamental---that represent $N_a$ copies of the tautological bundle. It would be interesting to test the non-perturbative dualities of \cite{Hori:2006dk, Benini:2012ui, Benini:2014mia, Gomis:2014eya}.

\subsection{Four-dimensional SQCD and Seiberg duality}

We consider a simple example of Seiberg duality for SQCD in four dimensions. The simplest model to study is $USp(2)$ SQCD with $2N_f =6$ flavors, whose global symmetry is $SU(6) \times U(1)_R$. The magnetic dual is a Wess-Zumino model of fifteen chiral multiplets, transforming as the antisymmetric tensor $M_{ij}$ ($i,j=1,\ldots, 6$) of $SU(6)$, interacting through the cubic superpotential $W = \Lambda^{-3} \,\text{Pf}\, M$ \cite{Intriligator:1995ne}. The $M_{ij}$ correspond to the mesons of the electric theory. We now check, at lowest order in $q$, that the partition functions of the two theories on $\Sigma_g\times T^2$ coincide.

Consider first the $USp(2)$ model. We need to satisfy the quantization condition (\ref{quantization R-charges}) for the R-charges. The quarks have R-charge $r=\frac13$ and the gauge invariants are the mesons $M_{ij}$, therefore for $g-1 \in 3\bZ$ the R-charges are correctly quantized. On the other hand, for generic $g$ we need to mix the exact R-symmetry of the IR fixed point with some flavor symmetry. As in \cite{Benini:2015noa}, we choose the non-anomalous and integer  R-symmetry $U(1)_R' = \text{diag}(1,1,0,0,0,0)$, which is a combination of the exact R-symmetry and an Abelian subgroup of the $SU(6)$ flavor symmetry. The latter is then broken to $SU(4)\times SU(2)\times U(1)$. For simplicity, we also choose a magnetic flavor flux along $U(1)_A = \text{diag}(-2,-2,1,1,1,1)$, thus preserving the residual flavor symmetry. We use the compact notation
\be
f_\chi(b,a,r) = \bigg( \frac{i\eta(q)}{ \theta_1(x^b y^a;q)} \bigg)^{b\fm + a\fn +(g-1)(r-1)} \;.
\ee
Then the classical and one-loop contribution to the partition function $Z_{\Sigma_g \times T^2}$ is
\be
Z_\text{cl,1l} = (i\, du) \, \eta(q)^{2(1-g)} \; \bigg( \frac{\theta_1(x^2;q)}{i\eta(q)} \; \frac{\theta_1(x^{-2};q)}{i\eta(q)} \bigg )^{1-g} \; \prod_{b=\pm1} f_\chi(b,-2,1)^2 \, f_\chi(b,1,0)^4 \;.
\ee
Formally the partition function is given by
\be
\label{Z 4d SU(2)}
Z_{\Sigma_g \times T^2} = \frac12 \sum_{\fm\in\bZ} \frac1{2\pi i} \int_\text{JK} Z_\text{cl,1l}  \, \bigg( \frac{\partial^2 \log Z_\text{cl,1l}}{i\partial u \, \partial \fm } \bigg)^g \;,
\ee
however in order to correctly take into account the contribution from W-bosons we should first sum over $\fm$ and then take the residues, avoiding the poles at the roots of the BAE. Choosing $\eta>0$ we need to collect the poles at $x=y^2$ and $x=y^{-1}$, and this dictates that we should resum over $\fm \geq -M$ for some large positive integer $M$. The resulting BAE is
\be
e^{iB} = \frac{\theta_1(x^{-1} y^{-2}; q)^2 \, \theta_1(x^{-1} y;q)^4 }{ \theta_1 (x y^{-2}; q)^2 \, \theta_1(xy; q)^4} = 1 \;,
\ee
to be solved on the torus $x \cong qx$.

Let us compute the partition function at the lowest order in $q$, in the limit $q\to 0$. Then the relevant solutions to the BAE are at the two roots of the polynomial equation
$$
2y^2(1+x^2) - x (1+2y -2y^2+2y^3+y^4) = 0 \;.
$$
We do not take the poles at $x=\pm1$ where the 1-loop gauge determinant vanishes, nor at $x=0$ which is outside the domain. We expand $Z_\text{cl,1l}$ at the lowest order in $q$ and apply (\ref{index formula schematic -- init})-(\ref{index formula schematic -- fin}). The result is
\be
Z_{\Sigma_g \times T^2} = - q^{5(g-1)/12} \bigg (y^{4(1-g)} (1+y)^{-8\fn} (1-y^2)^{-5(1-g)} (1+y^2)^{4\fn+1-g} + \cO(q) \bigg ) \;.
\ee
It is easy to check that this is precisely the expansion of the partition function of the dual theory
\be
Z_\text{dual} = f_\chi(0,-4,2) \, f_\chi(0,-1,1)^8 \, f_\chi(0,2,0)^6 \;,
\ee
up to a factor $(-1)^g$. With some effort, the analysis can be similarly extended to higher order in $q$, again with perfect agreement. It would clearly be desirable a proof of the equality for generic $q$, based on identities of theta functions. We leave such an analysis for future work.


\section{Large \matht{N} limit and black hole entropy}
\label{sec: entropy}

The topologically twisted index of the three-dimensional ABJM theory \cite{Aharony:2008ug} for $g=0$ has been computed in \cite{Benini:2015noa}, and its large $N$ limit matched with the Bekenstein-Hawking entropy of a class of AdS$_4$ supersymmetric black holes with horizon AdS$_2\times S^2$ and an embedding into M-theory \cite{Cacciatori:2009iz,Hristov:2010ri,Dall'Agata:2010gj}.  Similar black holes exist  with horizon AdS$_2\times \Sigma_g$.  Here we evaluate the index for generic $g$ and show that the matching with the entropy extends to arbitrary genus.

In $\cN=2$ notation, the  ABJM theory is a $U(N)_k\times U(N)_{-k}$ supersymmetric three-dimensional Chern-Simons theory (the subscripts are the CS levels) with  bi-fundamental chiral multiplets $A_i$ and $B_j$, $i,j=1,2$,  transforming in the $(N,\overline N)$ and $(\overline N, N)$ representations of the gauge group, respectively,  and subject to the superpotential
\be
\label{superpot}
W = \Tr \big( A_1 B_1 A_2 B_2 - A_1B_2 A_2 B_1 \big) \;.
\ee
We focus on $k=1$ where the theory has $\cN=8$ superconformal symmetry and $SO(8)$ \mbox{R-symmetry} and it is dual to AdS$_4\times S^7$. From the point of view of an $\cN=2$ subalgebra, the flavor symmetry appears to be $SU(2)\times SU(2)\times U(1)$ and its three Cartan generators give charges $(1,0,0,-1)$, $(0,1,0,-1)$, $(0,0,1,-1)$ to $(A_1, A_2, B_1, B_2)$ respectively, while we take R-charges $(0,0,0,2)$. We introduce fugacities $y_{1,2,3}$ and fluxes $-\fn_{1,2,3}$ for the three Cartan flavor symmetries of ABJM. The index can be written as
\be
\label{initial ZZ}
Z_g = \frac1{(N!)^2} \sum_{\fm, \wt\fm \,\in\, \bZ^N} \int_\text{JK} \;   Z_\text{cl,1l}  \, \bigg ( \det_{AB} \frac{\partial^2 Z_\text{cl,1l}}{i \, \partial u_A \, \partial \mathfrak{m}_B}\bigg )^g
\ee
where%
\footnote{As in \cite{Benini:2015noa}, we chose a convenient parameterization for the set of independent fugacities and fluxes. Topological symmetries have been identified
with a combination of flavor and gauge symmetries.}
\begin{multline}
\label{initial Z}
 Z_\text{cl,1l}  = \prod_{i=1}^N \frac{dx_i}{2\pi i x_i} \, \frac{d\tilde x_i}{2\pi i \tilde x_i} \, x_i^{\fm_i} \, \tilde x_i^{- \wt\fm_i} \times
\prod_{i\neq j}^N \Big( 1 - \frac{x_i}{x_j} \Big)^{1-g} \, \Big( 1 - \frac{\tilde x_i}{\tilde x_j} \Big)^{1-g} \times \\
\times \prod_{i,j=1}^N \prod_{a=1,2}
\bigg( \frac{ \sqrt{ \frac{x_i}{\tilde x_j} \, y_a} }{ 1- \frac{x_i}{\tilde x_j} \, y_a } \bigg)^{\fm_i - \wt\fm_j - \fn_a +1-g}
\prod _{b=3,4} \bigg( \frac{ \sqrt{ \frac{\tilde x_j}{x_i} \, y_b} }{ 1- \frac{\tilde x_j}{x_i} \, y_b } \bigg)^{\wt\fm_j - \fm_i - \fn_b +1-g} \;,
\end{multline}
and $u_A=(u_i,\tilde u_j), \mathfrak{m}_A=(\fm_i,\wt\fm_j)$ and $A,B=(i,j)$. In this formula $a=1,2$ refers to the fields $A_1$ and $A_2$ and $b=3,4$ to $B_1$ and $B_2$. The $\fn_a$ should be integers by the quantization condition:
\be
\fn_a \in \bZ \;.
\ee
In the expression above we have introduced the quantities $y_4$ and $\fn_4$ fixed by
\be
\label{quantization constraint}
\prod_{a=1}^4 y_a = 1 \;,\qquad\qquad \sum_{a=1}^4 \fn_a = 2 (1-g) \;.
\ee
They are useful since the results will be manifestly invariant under the $S_4$ that permutes the $U(1)$ factors in the Cartan of $SO(8)$.

Choosing covectors $-\eta = \wt\eta = (1,\ldots ,1)$, we can resum the integrand of \eqref{initial Z} with a cut-off  $\fm_i \leq M -1$ and $\wt\fm_j \geq -M$ for some large integer $M$. We obtain the very same BAE as in \cite{Benini:2015noa}, since the BAE does not depend on $g$:
\be
\label{BA expressions}
e^{iB_i} = x_i^k \prod_{j=1}^N \frac{ \big( 1- y_3 \frac{\tilde x_j}{x_i} \big) \big( 1- y_4 \frac{\tilde x_j}{x_i} \big) }{ \big( 1- y_1^{-1} \frac{\tilde x_j}{x_i} \big) \big( 1- y_2^{-1} \frac{\tilde x_j}{x_i} \big) } \;,\qquad
e^{i\wt B_j} = \tilde x_j^k \prod_{i=1}^N \frac{ \big( 1- y_3 \frac{\tilde x_j}{x_i} \big) \big( 1- y_4 \frac{\tilde x_j}{x_i} \big) }{ \big( 1- y_1^{-1} \frac{\tilde x_j}{x_i} \big) \big( 1- y_2^{-1} \frac{\tilde x_j}{x_i} \big) } \;.
\ee
The index is then obtained from \eqref{index formula schematic -- init}. In the large $N$ limit we expect that the sum in \eqref{index formula schematic -- init} will be dominated by the contribution of a single distribution $x_{(i)}$, $\tilde x_{(i)}$.

The large $N$ solution to \eqref{BA expressions} for real chemical potentials $y_a= e^{i\Delta_a}$ has been analyzed in detail in \cite{Benini:2015noa}. The large $N$ saddle-point eigenvalue distribution was found to be of the form
\be
\label{ansatz alpha}
u_i = i N^\frac12 t_i + v_i \;,\qquad\qquad \tilde u_i = i N^\frac12 t_i + \tilde v_i  \;.
\ee
In the large $N$ limit one defines the continuous functions $t(i/N) = t_i$, $v(i/N) = v_i$, $\tilde v(i/N) = \tilde v_i$ and  introduces the density of eigenvalues $ \rho(t) = N^{-1}  di/dt$, normalized as $ \int dt\, \rho(t) = 1$. The solution for $\rho(t)$ and $\delta v(t) = v(t)-\tilde v(t)$  is of a characteristic piecewise form, divided into regions bounded by the transition points 
\be
\label{solution sum 2pi -- init}
t_\ll = - \frac\mu{\Delta_3} \;,\qquad\quad t_< = - \frac\mu{\Delta_4} \;,\qquad\quad t_> = \frac\mu{\Delta_2} \;,\qquad\quad t_\gg = \frac\mu{\Delta_1} \;,
\ee
where we have chosen $\Delta_1 \leq \Delta_2$ and $\Delta_3 \leq \Delta_4$ without loss of generality. We are also taking $0 < \Delta_a < 2\pi$ and $\sum_a \Delta_a = 2\pi$. In the \emph{left tail} we have
\be
\begin{aligned}
\rho &= \frac{\mu + t\Delta_3}{(\Delta_1 + \Delta_3)(\Delta_2 + \Delta_3)(\Delta_4 - \Delta_3)} \\[.5em]
\delta v &= - \Delta_3 + e^{-\sqrt{N} Y_3} \;,\qquad\qquad Y_3 = \frac{- t\Delta_4 -\mu}{\Delta_4 - \Delta_3}
\end{aligned}
\qquad\qquad\qquad t_\ll < t < t_< \;.
\ee
In the \emph{inner interval} we have
\be
\begin{aligned}
\rho &= \frac{2\pi \mu + t(\Delta_3 \Delta_4 - \Delta_1 \Delta_2)}{(\Delta_1 + \Delta_3)(\Delta_2 + \Delta_3)(\Delta_1 + \Delta_4)(\Delta_2 + \Delta_4)} \\[.5em]
\delta v &= \frac{\mu(\Delta_1 \Delta_2 - \Delta_3 \Delta_4) + t \sum_{a<b<c} \Delta_a \Delta_b \Delta_c }{ 2\pi \mu + t ( \Delta_3 \Delta_4 - \Delta_1 \Delta_2) }
\end{aligned}
\qquad\qquad t_< < t < t_>
\ee
and $\delta v'>0$. In the \emph{right tail} we have
\be
\begin{aligned}
\rho &= \frac{\mu - t \Delta_1}{(\Delta_1 + \Delta_3)(\Delta_1 + \Delta_4)(\Delta_2 - \Delta_1)} \\[.5em]
\delta v &= \Delta_1 - e^{-\sqrt{N} Y_1} \;,\qquad\qquad Y_1 = \frac{t\Delta_2 - \mu}{\Delta_2 - \Delta_1}
\end{aligned}
\qquad\qquad\qquad t_> < t < t_\gg \;.
\ee
The normalization factor is $\mu = \sqrt{ 2 \Delta_1 \Delta_2 \Delta_3 \Delta_4}$. Notice that in the tails $\delta v$ is constant up to exponentially vanishing contributions in $N$, which are nevertheless important for the evaluation of the index.
 
The index is then computed by evaluating the contribution of the saddle-point solution to the sum  \eqref{index formula schematic -- init}.  The computation is essentially the same as in section 2.4 of \cite{Benini:2015noa} to which we refer for  details. The final result is
\begin{align}
\quad\rule{0pt}{2em} \re \log Z_g(\fn) &= - N^\frac32 \int dt\, \rho(t)^2 \bigg[ (1-g) \frac{2\pi^2}3 + \sum_{\substack{ a=3,4 \;:\; + \\ a=1,2 \;:\; -}} \big( \fn_a-1 + g \big)\, g_\pm'\big( \delta v(t) \pm \Delta_a \big) \bigg] \quad \nn \\
\rule[-2em]{0pt}{1em} &\quad - N^\frac32 \sum_{a=1}^4 \fn_a \int_{\delta v \,\approx\, \varepsilon_a \Delta_a} \hspace{-2em} dt\, \rho(t) \, Y_a(t) \quad \;,
\label{Z large N functional}
\end{align}
up to corrections of order $N\log N$. Here $g_\pm'(u) = \frac{u^2}2 \mp \pi u + \frac{\pi^2}3$. The first contribution in the first line of \eqref{Z large N functional} comes from the Vandermonde determinant and the second one from the matter contribution. The second line in \eqref{Z large N functional} comes from the matter contribution and the fermionic zero-modes on $\Sigma_g$. Since the logarithm of the one loop determinant of the chiral fields is singular on the tail regions, we need to take into account the exponentially small corrections $Y_a$ to the tails. The exponent $-\fn_a +1-g$ of the one-loop determinant is corrected to $-\fn_a$ by an analogous contribution from $\det(\partial B / \partial u)$ in \eqref{index formula schematic -- init}. 

By plugging $\rho$ and $\delta v$ into (\ref{Z large N functional}) we find
\be
\label{Z large N}
\re\log Z_g(\fn) = - \frac{N^\frac32}3 \sqrt{2\Delta_1 \Delta_2 \Delta_3 \Delta_4} \sum_{a=1}^4 \frac{\fn_a}{\Delta_a} = - N^\frac32 \frac{2\sqrt2}3 \sum_{a=1}^4 \fn_a \parfrac{}{\Delta_a} \sqrt{\Delta_1 \Delta_2 \Delta_3 \Delta_4} \;.
\ee
Indeed, it is obvious from (\ref{Z large N functional}) that $\re \log Z_g (\fn)  = (1-g) \re \log Z_{g=0}  \big( \fn / (1-g) \big)$, at least for $g\neq1$, and so the result for generic $g$ follows from the one for $g=0$ found in \cite{Benini:2015noa}. The expression in (\ref{Z large N}) looks independent from $g$, however one should recall that $\fn_a$ obey (\ref{quantization constraint}). The field theory entropy is obtained by extremizing \eqref{Z large N} with respect to $\Delta_{a=1,2,3}$ with $\sum_a \Delta_a = 2\pi$.
Other ${\cal N}=2$ theories could be similarly analized at higher genus, generalizing the results in \cite{Hosseini:2016tor, Hosseini:2016ume}.

\

We now compare the result with the  entropy of magnetically-charged BPS black holes (from M-theory on $S^7$) with horizon AdS$_2\times \Sigma_g$ and asymptotic to AdS$_4$. We  refer to appendices A and C of \cite{Benini:2015noa} for a detailed discussion of these black holes. The metric of the  black holes is of the form 
\be
ds^2 = - e^{2U(r)} dt^2 + e^{-2U(r)} \Big( dr^2 + e^{2V(r)} ds^2_{\Sigma_g} \Big) \;,
\ee
with scalar fields depending on the  radial coordinate. We choose a constant curvature metric $ds_\Sigma^2 = e^{2h}(dx^2 + dy^2)$ on the Riemann surface
with
\be
e^{2h} =
\begin{cases} \frac4{(1+x^2+y^2)^2} &\text{for } S^2 \\ 2\pi & \text{for } T^2 \\ \frac1{y^2} &\text{for } H^2 \end{cases} \quad .
\ee
 We have $R_s = 2\kappa$ where $\kappa = 1$ for $S^2$, $\kappa = 0$ for $T^2$, and $\kappa = -1$ for $H^2$. The range of coordinates are $(x,y) \in \bR^2$ for $S^2$, $(x,y) \in [0,1)^2$ for $T^2$, and $(x,y) \in \bR \times \bR_{>0}$ for $H^2$. In the $H^2$ case the upper half-plane has to be quotiented by a suitable Fuchsian group to get a compact Riemann surface $\Sigma_{g>1}$. The ranges are chosen in such a way that
\be
\Vol(\Sigma_g) = \int e^{2h} dx\,dy = 2\pi \eta \;,\qquad\qquad \eta \equiv \begin{cases} 2|g-1| &\text{for } g \neq 1 \\ 1 & \text{for } g = 1 \end{cases}
\ee
where we defined the positive number $\eta$.

The explicit expression for the entropy of the black holes is given in appendix A of \cite{Benini:2015noa}. The expression for the horizon area (proportional to the Bekenstein-Hawking entropy) is
\be
A = e^{2f_2 (\hat \fn)} \, 2\pi \eta \;,
\ee
in terms of magnetic charges $\hat \fn_a$.%
\footnote{The magnetic charges $\fn_a$ used in \cite{Benini:2015noa} are now called $\hat\fn_a$ to avoid confusion.}
The function $e^{2f_2}$, equal to $e^{2V-2U}$ at the horizon, is homogeneous under a common positive rescaling of the $\hat \fn_a$, with weight $1$. The charges $\hat \fn_a$ obey the BPS constraint $\sum_a \hat \fn_a = 2\kappa$ and the quantization condition $\hat\fn_a \in \frac2\eta \bZ$. It follows that we can identify
\be
\hat \fn_a = \frac2\eta \fn_a
\ee
with the integer magnetic charges $\fn_a$ used here. Then the horizon area, in terms of the $\fn_a$, is
\be
A = 4\pi \, e^{2f_2 (\fn)} \;,
\ee
with no explicit dependence on $g$. Therefore, the matching of the field theory index with the black hole entropy for $g=0$ exhibited in \cite{Benini:2015noa} implies matching for all values of $g$. The extremization of the quantity \eqref{Z large N} is the field theory counterpart of the attractor mechanism for AdS$_4$ black holes \cite{Dall'Agata:2010gj}.

\section*{Acknowledgements}

We thank Giulio Bonelli and Wenbin Yan for useful discussions.
FB is supported in part by the MIUR-SIR grant RBSI1471GJ ``Quantum Field Theories at Strong Coupling: Exact Computations and Applications'', by the INFN, and by the Royal Society as a Royal Society University Research Fellowship holder. AZ is supported by the INFN and the MIUR-FIRB grant RBFR10QS5J ``String Theory and Fundamental Interactions''.

\appendix

\section{Notation, Lagrangians and supersymmetry variations}
\label{app: notation}

The 3d $\cN=2$ vector multiplet $\cV$ contains the fields $(A_\mu, \sigma, \lambda, \lambda^\dag, D)$: they are the gauge field, a real scalar, a Dirac spinor and its conjugate, and the auxiliary real scalar, respectively, all in the adjoint representation of the gauge group $G$. In Euclidean signature they get complexified. They have R-charges $(0,0,-1,1,0)$ respectively. The supersymmetry variations, specialized to constant positive-chirality spinors $\epsilon$, $\tilde\epsilon$, are:%
\footnote{The transformation are valid for generic covariantly constant spinors $\epsilon$, $\tilde\epsilon$.}
\bea
\label{vector variations}
Q A_\mu &= \frac i2 \lambda^\dag \gamma_\mu \epsilon &
Q\lambda &= + \frac12 \gamma^{\mu\nu} \epsilon F_{\mu\nu} - D\epsilon + i \gamma^\mu \epsilon \, D_\mu\sigma \\
\wt Q A_\mu &= \frac i2 \tilde\epsilon^\dag \gamma_\mu \lambda &
\wt Q \lambda^\dag &= - \frac12 \tilde\epsilon^\dag \gamma^{\mu\nu} F_{\mu\nu} + \tilde\epsilon^\dag D + i \tilde\epsilon^\dag \gamma^\mu D_\mu\sigma \\
QD &= - \frac i2 D_\mu \lambda^\dag \gamma^\mu \epsilon + \frac i2 [\lambda^\dag \epsilon, \sigma ] \qquad\quad &
Q \lambda^\dag &= 0 \qquad\qquad\qquad
Q\sigma = - \frac12 \lambda^\dag \epsilon \\
\wt Q D &= \frac i2 \tilde\epsilon^\dag \gamma^\mu D_\mu \lambda + \frac i2 [\sigma, \tilde\epsilon^\dag\lambda] &
\wt Q \lambda &= 0 \qquad\qquad\qquad
\wt Q \sigma = - \frac12 \tilde\epsilon^\dag \lambda
\eea
and then
\be
Q F_{\mu\nu} = i D_{[\mu} \lambda^\dag \gamma_{\nu]} \epsilon \;,\qquad\qquad \wt Q F_{\mu\nu} = - i \tilde\epsilon^\dag \gamma_{[\mu} D_{\nu]} \lambda \;.
\ee

The chiral multiplet $\Phi$ contains the fields $(\phi, \psi, F)$: a complex scalar, a Dirac spinor and the auxiliary complex scalar, in a representation $\fR$ of $G$. They have R-charges $(q, q-1, q-2)$. The anti-chiral multiplet $\wb\Phi$ contains $(\phi^\dag, \psi^\dag, F^\dag)$ in representation $\wb\fR$ and with opposite R-charges. The supersymmetry variations are
\bea
Q\phi &= 0 \qquad\quad & Q\psi &= \big( i\gamma^\mu D_\mu\phi + i \sigma\phi \big)\epsilon & \wt Q\psi &= \tilde\epsilon^c F \\
\wt Q\phi &= - \tilde\epsilon^\dag\psi \qquad & \wt Q\psi^\dag &= \tilde\epsilon^\dag \big( -i\gamma^\mu D_\mu \phi^\dag + i \phi^\dag \sigma \big) & Q\psi^\dag &= - \epsilon^{c\dag} F^\dag \\
Q\phi^\dag &= \psi^\dag\epsilon \qquad & QF &= \epsilon^{c\dag} \big( i \gamma^\mu D_\mu \psi - i \sigma\psi - i \lambda\phi \big) & \wt QF &= 0 \\
\wt Q\phi^\dag &= 0 \qquad & \wt Q F^\dag &= \big( -i D_\mu\psi^\dag \gamma^\mu - i \psi^\dag \sigma + i \phi^\dag \lambda^\dag \big) \tilde\epsilon^c \qquad\quad & QF^\dag &= 0 \;.
\eea
We define charge conjugate spinors $\epsilon^c = C \epsilon^*$ and  $\epsilon^{c\dag} = \epsilon^\trans C$, where $C$ is the charge conjugation matrix such that $C\gamma^\mu C^{-1} = - \gamma^{\mu\trans}$. We choose $C = \gamma_2$ so that $C = C^{-1} = C^\dag = - C^\trans = - C^*$.

The Lagrangian terms are essentially equal to those on flat space, with the only difference that they are coupled to the R-symmetry and flavor background. The Yang-Mills action is
\be
\cL_\text{YM} = \Tr\bigg[ \frac14 F_{\mu\nu} F^{\mu\nu} + \frac12 D_\mu\sigma D^\mu\sigma + \frac12 D^2 - \frac i2 \lambda^\dag \gamma^\mu D_\mu\lambda - \frac i2 \lambda^\dag [\sigma,\lambda] \bigg] \;.
\ee
The Chern-Simons Lagrangian for each simple factor $G_I$ in $G$ is
\be
\cL_\text{CS}^\text{n-Ab} = - \frac{ik_I}{4\pi} \Tr \bigg[ \epsilon^{\mu\nu\rho} \Big( A_\mu \partial_\nu A_\rho - \frac{2i}3 A_\mu A_\nu A_\rho \Big) + \lambda^\dag \lambda + 2D\sigma \bigg] \;.
\ee
In particular there is a separate integer Chern-Simons coupling $k_I$ for each simple factor.
The Chern-Simons Lagrangian for the Abelian factors is
\be
\cL_\text{CS}^\text{Ab} = - \frac{ik_{ij}}{4\pi} \bigg[ \epsilon^{\mu\nu\rho} A_\mu^{(i)} \partial_\nu A_\rho^{(j)} + \lambda^{(i)\dag} \lambda^{(j)} + 2D^{(i)}\sigma^{(j)} \bigg] \;,
\ee
where $k_{ij}$ is a symmetric integer matrix. The kinetic Lagrangian for chiral multiplets is
\be
\cL_\text{mat} = D_\mu \phi^\dag D^\mu\phi + \phi^\dag \big( \sigma^2 + iD - q_\phi W_{12} \big) \phi + F^\dag F + i \psi^\dag ( \gamma^\mu D_\mu -\sigma ) \psi - i \psi^\dag \lambda\phi + i \phi^\dag \lambda^\dag \psi \;.
\ee
Notice that $W_{12}$ couples to each field $\phi$ proportionally to its R-charge $q_\phi$.
Superpotential interactions are described by the Lagrangians
\be
\cL_W = i F_W \;,\qquad\qquad\qquad \cL_{\wb W} = i F_W^\dag \;,
\ee
where
\be
F_W = \parfrac{W}{\Phi_i} F_i - \frac12\, \parfrac{^2W}{\Phi_i \partial\Phi_j} \psi_j^{c\dag} \psi_i\pdag \;,\qquad\qquad F_W^\dag = \parfrac{\wb W}{\Phi_i^\dag} F_i^\dag - \frac12\, \parfrac{^2 \wb W}{\Phi_i^\dag \partial\Phi_j^\dag} \psi_j^\dag \psi_i^c
\ee
are the F-terms of the chiral multiplet $W(\Phi)$ and its antichiral partner, separately supersymmetric.

\bibliographystyle{JHEP}
\bibliography{S2A}

\end{document}